\documentclass{article}

\PassOptionsToPackage{numbers, compress, sort}{natbib}

\usepackage{graphicx}
\usepackage{subcaption}
\usepackage{amsmath}
\usepackage{tikz}
\usepackage{multirow}
\usepackage{tabularx}
\usepackage[colorinlistoftodos]{todonotes}
\usetikzlibrary{automata, positioning, calc, shapes.multipart, fit, shapes.geometric}
\usetikzlibrary{decorations.text,decorations.pathmorphing,decorations.pathmorphing, decorations.pathreplacing, decorations.shapes}
\usepackage{bm}
\usepackage{siunitx}

\usepackage[final]{neurips_2024}

\usepackage[utf8]{inputenc} % allow utf-8 input
\usepackage[T1]{fontenc}    % use 8-bit T1 fonts
\usepackage{hyperref}       % hyperlinks
\usepackage{url}            % simple URL typesetting
\usepackage{booktabs}       % professional-quality tables
\usepackage{amsfonts}       % blackboard math symbols
\usepackage{nicefrac}       % compact symbols for 1/2, etc.
\usepackage{microtype}      % microtypography
\usepackage{xcolor}         % colors
\usepackage{booktabs}

\usepackage{listings}
\usepackage{xcolor}

\usepackage{textcomp}

\definecolor{codebackground}{rgb}{1,1,1} % white
\definecolor{commentcolor}{RGB}{0,128,0} % green
\definecolor{keywordcolor}{RGB}{0,0,255} % blue
\definecolor{numbercolor}{RGB}{128,128,128} % grey
\definecolor{stringcolor}{RGB}{160,32,240} % purple

\lstdefinestyle{verilogstyle}{
    backgroundcolor=\color{codebackground},   
    commentstyle=\color{commentcolor},
    keywordstyle=\color{keywordcolor},
    numberstyle=\color{numbercolor},
    stringstyle=\color{stringcolor},
    basicstyle=\ttfamily\scriptsize,
    breakatwhitespace=false,
    breaklines=true,
    captionpos=b,
    keepspaces=true,
    numbers=left,
    numbersep=5pt,
    showspaces=false,
    showstringspaces=false,
    showtabs=false,
    tabsize=4,
    language=Verilog
}

\DeclareMathOperator{\ltlnext}{\sf X}
\DeclareMathOperator{\ltlglobally}{\sf G}
\DeclareMathOperator{\ltlfinally}{\sf F}
\DeclareMathOperator{\ltluntil}{\sf U}
\DeclareMathOperator{\ltlFG}{\sf FG}
\DeclareMathOperator{\ltlGF}{\sf GF}
\DeclareMathOperator{\relu}{ReLU}
\DeclareMathOperator{\reg}{reg}
\DeclareMathOperator{\obs}{obs}
\DeclareMathOperator{\inp}{inp}
\newcommand{\signal}[1]{\mbox{\tt #1}}
\def\bq{\boldsymbol{q}}
\def\bs{\boldsymbol{s}}
\def\br{\boldsymbol{r}}
\DeclareMathOperator{\clamp}{Clamp}
\def\fr{f}

\title{Neural Model Checking}

\author{%
  Mirco Giacobbe\thanks{The authors are listed alphabetically.}
  \\
  University of Birmingham, UK
   \And
    Daniel Kroening\footnotemark[1]\\
  Amazon Web Services, USA
   \AND
   Abhinandan Pal\footnotemark[1]\\
  University of Birmingham, UK 
  \And
   Michael Tautschnig\footnotemark[1]\\
  Amazon Web Services, USA and\\ Queen Mary University of London, UK
}

\begin{document}

\maketitle

\begin{abstract}
  We introduce a machine learning approach to model checking temporal logic,
  with application to formal hardware verification.  Model checking answers
  the question of whether every execution of a given system satisfies a
  desired temporal logic specification.  Unlike testing, model checking
  provides formal guarantees.  Its application is expected standard in
  silicon design and the EDA industry has invested decades into the
  development of performant symbolic model checking algorithms.  Our new
  approach combines machine learning and symbolic reasoning by using neural
  networks as formal proof certificates for linear temporal logic.  We train
  our neural certificates from randomly generated executions of the system
  and we then symbolically check their validity using satisfiability solving
  which, upon the affirmative answer, establishes that the system provably
  satisfies the specification.  We leverage the expressive power of neural
  networks to represent proof certificates as well as the fact that checking
  a certificate is much simpler than finding one.  As a result, our machine
  learning procedure for model checking is entirely unsupervised, formally
  sound, and practically effective.  We experimentally demonstrate that our
  method outperforms the state-of-the-art academic and commercial model
  checkers on a set of standard hardware designs written in SystemVerilog.
\end{abstract}

\section{Introduction}
\label{sec:intro}

Electronic design is complex and prone to error.  Hardware bugs are
permanent after production and as such can irremediably affect the
correctness of software---which runs on hardware---and can compromise the
safety of cyber-physical systems---which embed hardware.  Correctness
assurance is core to the engineering of digital circuitry, with the median
FPGA and IC/ASIC projects spending respectively \qty{40}{\percent} and
\qty{60}{\percent} of time in verification~\cite{wilson}.  Verification
approaches based on directed or constrained random testing are easy to set
up but are inherently non-exhaustive~\cite{spearsystemverilog,
DBLP:conf/uss/TrippelSCKRH22}.  Testing cannot show the absence of bugs
which, for systems the safety of which is critical, can have serious
consequences; notably, over \qty{40}{\percent} of hardware development
projects must satisfy at least one functional safety standard~\cite{wilson}. 
In contrast to testing, {\em model checking} a design against a formal
specification of correctness answers the question of whether the design
satisfies the specification with mathematical certainty, for every possible
execution of the system~\cite{mc2nd, principlesofmc, principlesofcps}.

The EDA industry has heavily invested in software tools for symbolic model
checking.  Early symbolic model checking algorithms utilise fixed-point
computations with binary decision diagrams
(BDDs)~\cite{DBLP:journals/tc/Akers78}, where each node specifies the
Boolean assignment for a circuit's flip-flop or input
bit~\cite{DBLP:journals/iandc/BurchCMDH92, DBLP:conf/icalp/EmersonC80}. 
BDDs struggle to scale when applied to complex arithmetic data paths,
prompting a shift towards iterative approximation of fixed points using
propositional satisfiability (SAT) solving~\cite{DBLP:conf/tacas/BiereCCZ99,
DBLP:conf/cav/BiereCRZ99, DBLP:conf/tacas/ClarkeKL04}, which is now the
state-of-the-art technique.  Both BDD and SAT-based model checking, despite
extensive research, remain computationally demanding; even small circuit
modules can require days to verify or may not complete at all. 
Consequently, verification engineers often limit state space exploration to
a bounded time horizon through bounded model checking, sacrificing global
correctness over the unbounded time domain.

We present a machine learning approach to hardware model checking that
leverages neural networks to represent proof certificates for the compliance
of a given hardware design with a given linear temporal logic (LTL)
specification~\cite{DBLP:conf/focs/Pnueli77}.  Our approach avoids
fixed-point algorithms entirely, capitalises on the efficient word-level
reasoning of satisfiability solvers, and delivers a formal guarantee over an
unbounded time horizon.  Given a hardware design and an LTL specification
$\Phi$, we train a word-level neural certificate for the compliance of the
design with the specification from test executions, which we then check
using a satisfiability solver.  We leverage the observation that checking a
proof certificate is much simpler than solving the model checking problem
directly, and that neural networks are an effective representation of proof
certificates for the correctness of systems~\cite{DBLP:conf/nips/ChangRG19,
DBLP:conf/sigsoft/GiacobbeKP22}.  We ultimately obtain a machine learning
procedure for hardware model checking that is entirely unsupervised,
formally sound and, as our experiments show, very effective in practice.

Our learn-and-check procedure begins by generating a synthetic dataset
through random executions of the system alongside a B\"uchi automaton that
identifies counterexamples to $\Phi$.  We then train a \emph{neural ranking
function} designed to strictly decrease whenever the automaton encounters an
accepting state and remain stable on non-accepting states.  After training,
we formally check that the ranking function generalises to all possible
executions.  We frame the check as a cost-effective one-step bounded model
checking problem involving the system, the automaton, and the quantised
neural ranking function, which we delegate to a satisfiability solver.  As
the ranking function cannot decrease indefinitely, this confirms that the
automaton cannot accept any system execution, effectively proving that such
executions are impossible.  Hence, if the solver concludes that no
counterexample exists, it demonstrates that no execution satisfies $\lnot
\Phi$, thereby affirming that the system satisfies
$\Phi$~\cite{DBLP:journals/apal/Vardi91, DBLP:conf/popl/CookGPRV07}.

We have built a prototype that integrates PyTorch, the bounded model checker
EBMC, the LTL-to-automata translator Spot, the SystemVerilog simulator
Verilator, and the satisfiability solver
Bitwuzla~\cite{DBLP:conf/isvlsi/MukherjeeKM15, snyder2004verilator,
DBLP:conf/cav/NiemetzP23, DBLP:conf/cav/Duret-LutzRCRAS22}.  We have
assessed the effectiveness of our method across 194 standard hardware model
checking problems written in SystemVerilog and compared our results with the
state-of-the-art academic hardware model checkers ABC and
nuXmv~\cite{DBLP:conf/cav/BraytonM10, DBLP:conf/cav/CavadaCDGMMMRT14}, and
two commercial counterparts.  For any given time budget of less than 5
hours, our method completes on average \qty{60}{\percent} more tasks than ABC, \qty{34}{\percent}
more tasks than nuXmv, and \qty{11}{\percent} more tasks than the leading commercial
model checker.  
Our method is faster than the academic tools on
\qty{67}{\percent} of the tasks, 10X faster on \qty{34}{\percent}, and 100X
faster on \qty{4}{\percent}; when considering the leading commercial tool,
our method is faster on \qty{75}{\percent}, 10X faster on
\qty{29}{\percent}, and 100X faster on \qty{2}{\percent} of them.  Overall,
with a straightforward implementation, our method outperforms mature
academic and commercial model checkers.

Our contribution is threefold.  We present for the first time a hardware
model checking approach based on neural certificates.  We extend neural
ranking functions, previously introduced for the termination analysis of
software, to LTL model checking and the verification of reactive systems. 
We have built a prototype and experimentally demonstrated that our approach
compares favourably with the leading academic and commercial hardware model
checkers.  Our technology delivers formal guarantees of correctness and
positively contributes to the safety assurance of systems.

\section{Automata-theoretic Linear Temporal Logic Model Checking}
\label{sec:Automata}

An LTL model checking problem consists of a model ${\cal M}$ that describes
a system design and an LTL formula $\Phi$ that describes the desired
temporal behaviour of the system~\cite{DBLP:conf/focs/Pnueli77,
DBLP:conf/nato/HarelP84}.  The problem is to decide whether all traces of
${\cal M}$ satisfy $\Phi$.

Our formal model $\cal M$ of a hardware design consists of a finite set of
bit-vector-typed variables $X_{\cal M}$ with fixed bit-width and domain of
assignments~$S$, partitioned into input variables $\inp X_{\cal M} \subseteq
X_{\cal M}$ and state-holding register variables $\reg X_{\cal M} \subseteq
X_{\cal M}$; we interpret primed variables $X_{\cal M}'$ as the value of
$X_{\cal M}$ after one clock cycle.  Then, a sequential update relation
$\rm Update_{\cal M}$ relates $X_{\cal M}$ and $\reg X_{\cal M}'$ and computes
the next-state valuation of the registers from the current-state valuation
of all variables; we interpret $\rm Update_{\cal M}$ as a first-order logic
formula encoding this relation.  A state $\boldsymbol{s} \in S$ is a
valuation for the variables $X_{\cal M}$.  We denote as $\reg \bs, \inp \bs,
\dots $ the restriction of $\bs$ to the respective class of variables.  For
two states $\boldsymbol{s}$ and $\boldsymbol{s'}$, the state
$\boldsymbol{s}'$ is a successor of $\boldsymbol{s}$, which we write as
$\boldsymbol{s} \rightarrow_{\cal M} \boldsymbol{s}'$, if ${\rm
Update}_{\cal M}(\boldsymbol{s}, \reg \boldsymbol{s}')$ evaluates to true. 
We call $\rightarrow_{\cal M}$ the transition relation of $\cal M$ and say
that an infinite sequence of states $\bar{\bs}_0, \bar{\bs}_1, \bar{\bs}_2,
\dots$ is an execution of $\cal M$ if $\bar{\bs}_i \rightarrow_{\cal M}
\bar{\bs}_{i+1}$ for all $i \geq 0$; we say that an execution is initialised
in $s_0 \in S$ when $\bar{\bs}_0 = s_0$.

We specify the intended behaviour of systems in LTL, which is the foundation
of SystemVerilog Assertions.  LTL extends propositional logic with temporal
modalities $\ltlnext$, $\ltlglobally$, $\ltlfinally$, and $\ltluntil$.  The
modality $\ltlnext\,\Phi_1$ indicates that $\Phi_1$ holds immediately after
one step in the future, $\ltlglobally\,\Phi_1$ indicates that $\Phi_1$ holds
at all times in the future, $\ltlfinally\,\Phi_1$ indicates that $\Phi_1$
holds at some time in the future, and $\Phi_1\,\ltluntil\,\Phi_2$ indicates
that $\Phi_1$ holds at all times until $\Phi_2$ holds at some time in the
future.  We refer the reader to the literature for the formal syntax and
semantics of LTL~\cite{DBLP:conf/focs/Pnueli77}.  The atomic propositions of
the LTL formulae we consider are Boolean variables of $\cal M$, which we
call the observables $\obs X_{\cal M} \subseteq X_{\cal M}$ of~$\cal M$.  We
note that any first-order predicate over $X_{\cal M}$ can be bound to a
Boolean observable using combinational logic
(cf.~Figure~\ref{fig:illustrativeEx}, where observable {\tt ful} corresponds
to predicate {\tt cnt == 7}).

We call a trace of $\cal M$ a sequence $\obs \bar{\bs}_0, \obs \bar{\bs}_1,
\obs \bar{\bs}_2, \dots$ where $\bar{\bs}_0, \bar{\bs}_1, \bar{\bs}_2,
\dots$ is an execution of~$\cal M$.  We define the language $L_{\cal M}$ of
$\cal M$ as the maximal set of traces of $\cal M$.  Every LTL formula~$\Phi$
is interpreted over traces and as such defines the language $L_{\Phi}$ of
traces that satisfy $\Phi$.  The model checking problem corresponds to
deciding the language inclusion question $L_{\cal M} \subseteq L_\Phi$.

As is standard in automata-theoretic model checking, we rely on the result
that every LTL formula admits a non-deterministic B\"uchi automaton that
recognises the same language~\cite{DBLP:journals/apal/Vardi91,
DBLP:conf/lics/VardiW86}.  A non-deterministic B\"uchi automaton $\cal A$
consists of a finite set of states $Q$, an initial start state $q_0 \in Q$,
an input domain $\Sigma$ (also called alphabet), a transition relation
$\delta \subseteq Q \times \Sigma \times Q$, and a set of fair states $F
\subseteq Q$.  One can interpret an automaton $\cal A$ as a hardware design
with one register variable $\reg X_{\cal A} = \{\text{q}\}$ having domain
$Q$, input and observable variables $\inp X_{\cal A} = \obs X_{\cal A}$
having domain $\Sigma$, and sequential update relation ${\rm Update}_{\cal
A}(\boldsymbol{\sigma}, \bq, \bq') \equiv (\bq, \boldsymbol{\sigma}, \bq')
\in \delta$ governing the automaton state transitions.  We say that an
execution of $\cal A$ is \emph{fair} (also said to be an accepting
execution) if it visits fair states infinitely often.  We define the fair
language $L^{\rm f}_{\cal A}$ of $\cal A$ as the maximal set of traces
corresponding to fair executions initialised in $q_0$.  Given any LTL
formula $\Phi$, there are translation algorithms and tools to construct
non-deterministic B\"uchi automata ${\cal A}_\Phi$ such that $L^{\rm
f}_{{\cal A}_\Phi} = L_\Phi$~\cite{DBLP:conf/cav/Duret-LutzRCRAS22,
DBLP:conf/atva/KretinskyMS18}.

\begin{figure}
\centering
\begin{tabular}{cc}
\begin{tikzpicture}
    \def\nncolor{black}
    \def\nnsep{1.5pt}
    \def\nnyd{7pt}
    \def\nnxd{15pt}
    \def\nnx{0cm}
    \def\nny{7pt}
    \def\nnboxsep{5pt}
    \node[draw, circle, thick, inner sep=\nnsep,\nncolor] (i1) at ($(\nnx+\nnboxsep,\nny+\nnyd)$) {};
    \node[draw, circle, thick, inner sep=\nnsep,\nncolor] (i2) at ($(\nnx+\nnboxsep,\nny)$) {};
    \node[draw, circle, thick, inner sep=\nnsep,\nncolor] (i3) at ($(\nnx+\nnboxsep,\nny-\nnyd)$) {};
    \node[draw, circle, thick, inner sep=\nnsep,\nncolor] (h1) at ($(\nnx+\nnboxsep+\nnxd,\nny+1.5*\nnyd)$) {};
    \node[draw, circle, thick, inner sep=\nnsep,\nncolor] (h2) at ($(\nnx+\nnboxsep+\nnxd,\nny+.5*\nnyd)$) {};
    \node[draw, circle, thick, inner sep=\nnsep,\nncolor] (h3) at ($(\nnx+\nnboxsep+\nnxd,\nny-.5*\nnyd)$) {};
    \node[draw, circle, thick, inner sep=\nnsep,\nncolor] (h4) at ($(\nnx+\nnboxsep+\nnxd,\nny-1.5*\nnyd)$) {};
    \node[draw, circle, thick, inner sep=\nnsep,\nncolor] (o1) at ($(\nnx+\nnboxsep+2*\nnxd,\nny)$) {};
    \draw[\nncolor] (i1) -- (h1) -- (i2) -- (h2) -- (i3) -- (h3) -- (o1) -- (h1)
    -- (i3) -- (h4) -- (o1) -- (h2) -- (i1) -- (h3) -- (i2) -- (h4) -- (i1);

    \tikzstyle{box} = [minimum width=2*\nnxd+2*\nnboxsep, minimum height=3*\nnyd+2*\nnboxsep, draw, thick]
    \node[box] (Mod) at (-1.6cm,+\nnyd) {$\mathcal{M}$};
    \node[box] (Aut) at (-1.6cm,-15mm) {$\mathcal{A}_{\lnot \Phi}$};
    \node[box] (Net) at ($(\nnx+\nnboxsep+\nnxd, \nny)$) {};

    \node[trapezium, trapezium angle=60, minimum height=0.5*(3*\nnyd+2*\nnboxsep), draw, thick] at ($(\nnx+\nnboxsep+\nnxd, -15mm)$) (Par) {};

    \def\inpsh{5mm}
    \def\statesh{3mm}

    \path[thick] let \p1=($(Mod.east)$), \p2=($(Net.west)+(0,0)$), \p3=(Mod.north west), \p4=($(Mod.north west)!0.33!(Mod.south west)$) in 
    (\p1) edge[-] ($(0.5*\x1+0.5*\x2, \y1)$)
    ($(0.5*\x1+0.5*\x2, \y1)$) edge[-] ($(0.5*\x1+0.5*\x2, \y2)$) node[fill=black, circle, inner sep=1pt] {}
    ($(0.5*\x1+0.5*\x2, \y2)$) edge[->] (\p2)
    ($(0.5*\x1+0.5*\x2, \y2)$) edge[-] ($(0.5*\x1+0.5*\x2, \y3+\statesh)$)
    ($(0.5*\x1+0.5*\x2, \y3+\statesh)$) edge[-] node[above] {$\reg X_{\cal M}$}  ($(\x3-\inpsh, \y3+\statesh)$)
    ($(\x3-\inpsh, \y3+\statesh)$) edge[-] ($(\x3-\inpsh, \y4)$) 
    ($(\x3-\inpsh, \y4)$) edge[->] (\p4);
    \path[thick] let \p1=($(Aut.north east)!0.5!(Aut.south east)$), \p2=($(Net.west)-(0,\nnyd)$), \p3=(Aut.south west), \p4=($(Aut.north west)!0.5!(Aut.south west)$) in 
    (\p1) edge[-] ($(0.5*\x1+0.5*\x2, \y1)$)

    ($(0.5*\x1+0.5*\x2, \y1)$) node[fill=black, circle, inner sep=1pt] {} edge[-] ($(0.5*\x1+0.5*\x2, \y3-\statesh)$)
    ($(0.5*\x1+0.5*\x2, \y3-\statesh)$) edge[-] node[below] {q} ($(\x3-\inpsh, \y3-\statesh)$)
    ($(\x3-\inpsh, \y3-\statesh)$) edge[-] ($(\x4-\inpsh, \y4)$) 
    ($(\x4-\inpsh, \y4)$) edge[->] (\p4)
    ($(0.5*\x1+0.5*\x2, \y1)$) edge[->] (Par);
    \draw[thick] ($(Mod.south west)!0.66!(Mod.south east)$) edge[->] node[right]{$\obs X_{\cal M}$} ($(Aut.north west)!0.66!(Aut.north east)$);
    \draw[thick] let \p1=($(Mod.north west)!0.66!(Mod.south west)$) in
    ($(\p1)-(\inpsh,0)$) edge[->]  (\p1) node[left] {$\inp X_{\cal M}$};

    \path[thick] let \p1=($(Mod.south west)!0.33!(Mod.south east)$), \p2=($(Aut.north west)!0.33!(Aut.north east)$), \p3=($(Mod.west)$) in
    ($(\x3-\inpsh, 0.5*\y1 + 0.5*\y2)$) node[left]{clk} edge[-]  ($(\x1, 0.5*\y1 + 0.5*\y2)$) 
    ($(\x1, 0.5*\y1 + 0.5*\y2)$) node[fill=black, circle, inner sep=1pt] {} edge[->] (\p1)
    ($(\x1, 0.5*\y1 + 0.5*\y2)$) edge[->] (\p2);
    \path[thick] let \p1=($(Aut.north east)!0.33!(Aut.south east)$), \p2=(Net.east) in
    (\p2) edge[->] ($(\p2)+(\inpsh,0)$) ($(\p2)+(\inpsh,0)$) node[right] {$V$};
    \path[thick] ($(Par.south)-(0,5mm)$) node[below] {$\theta$} edge[->] (Par.south);
    \path[thick] (Par.north) edge[->] node[right] {$\theta_{\text{q}}$} (Net.south);
\end{tikzpicture} & \hspace{-6pt}\begin{tikzpicture}[scale=1]
  % Local parameters
  \def\plotwidth{30} % Width of the plot
  \def\plotheight{2.8} % Height of the plot
  \def\interval{5.5}
  \def\stepH{0.3}
  \def\offset{2.3}
  \def\stepHR{0.2}
  \def\syncdist{0}

  % Gridlines
  \foreach \x in {0,1,...,\plotwidth} {
    \draw [dotted] (\x/\interval,-\stepH/2+\syncdist) -- (\x/\interval,\plotheight+\syncdist);
  }

  \draw[black,thick] (0,\offset+\syncdist) -- (3/\interval,\offset+\syncdist) -- (3/\interval, \offset - \stepHR+\syncdist ) -- (4/\interval, \offset - \stepHR +\syncdist) -- (5/\interval, \offset - \stepHR +\syncdist) -- (5/\interval, \offset - 2*\stepHR +\syncdist) -- (6/\interval, \offset - 2*\stepHR +\syncdist) -- (8/\interval, \offset - 2*\stepHR +\syncdist) -- (8/\interval, \offset - 3*\stepHR +\syncdist)  -- (9/\interval, \offset - 3*\stepHR +\syncdist) -- (9/\interval, \offset - 4*\stepHR +\syncdist)-- (10/\interval, \offset - 4*\stepHR +\syncdist) -- (13/\interval, \offset - 4*\stepHR +\syncdist)  -- (13/\interval, \offset - 5*\stepHR +\syncdist) -- (16/\interval, \offset - 5*\stepHR +\syncdist) -- (16/\interval, \offset - 6*\stepHR +\syncdist) -- (30/\interval, \offset - 6*\stepHR +\syncdist);

  \draw[red,thick, dotted, line width=2.pt] (0,\offset - 6*\stepHR +\syncdist) -- (30/\interval, \offset - 6*\stepHR +\syncdist);
  
  % Waveform
  \draw[black,thick] (0,0+\syncdist) -- (2/\interval,0 +\syncdist) -- (2/\interval, \stepH+\syncdist) -- (3/\interval, \stepH+\syncdist) -- (3/\interval, 0+\syncdist) -- (4/\interval, 0+\syncdist) -- (4/\interval, \stepH+\syncdist) -- (5/\interval, \stepH+\syncdist) -- (5/\interval, 0+\syncdist) -- (7/\interval, 0+\syncdist) -- (7/\interval, \stepH+\syncdist) -- (9/\interval, \stepH+\syncdist) -- (9/\interval, 0+\syncdist) -- (12/\interval, 0+\syncdist) -- (12/\interval, \stepH+\syncdist)  -- (13/\interval, \stepH+\syncdist) -- (13/\interval, 0+\syncdist) -- (15/\interval, 0+\syncdist) -- (15/\interval, \stepH+\syncdist) -- (16/\interval, \stepH+\syncdist) -- (16/\interval, 0+\syncdist) -- (30/\interval, 0+\syncdist);

  \node at(-.5,0.2+\syncdist) {${\bf 1}_F(\text{q})$};
  \node at(-.5,\offset-0.2+\syncdist) {$V$};
  \node at(2.75,-0.5+\syncdist) {time $\rightarrow$};
  \draw[opacity=0] (0,0) circle (1);
\end{tikzpicture}\\
(a) & (b)
\end{tabular}
\caption{Automata-theoretic neural model checking via fair termination}
\label{fig:block-diagram}
\end{figure}
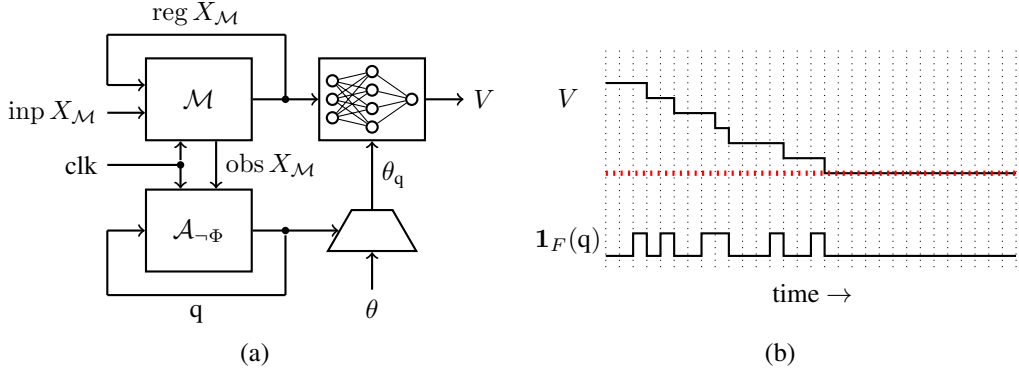

The standard approach to answer the language inclusion question $L_{\cal M} \subseteq L_{\Phi}$ is to answer the dual language emptiness question $L_{\cal M} \cap L_{\lnot \Phi} = \emptyset$~\cite{mc2nd,principlesofmc}. For this purpose, we first construct a non-deterministic B\"uchi automaton ${\cal A}_{\lnot \Phi}$ for the complement specification $\lnot \Phi$ where $\inp X_{{\cal A}_{\lnot \Phi}} = \obs X_{\cal M}$, then we reason over the synchronous composition (over a shared clock) of $\cal M$ and ${\cal A}_{\lnot \Phi}$ as illustrated in Figure~\ref{fig:block-diagram}a. We direct the reader to the relevant literature for general definitions of system composition~\cite{DBLP:conf/lics/AlurH96}. 
In this context, the synchronous composition results in the system ${\cal M} \parallel {{\cal A}_{\lnot \Phi}}$ with input variables $\inp X_{{\cal M} \parallel {{\cal A}_{\lnot \Phi}}} = \inp X_{\cal M}$, 
register variables $\reg X_{{\cal M} \parallel {{\cal A}_{\lnot \Phi}}} = \reg X_{\cal M} \cup \{ \text{q} \}$,
observable variables $\obs X_{{\cal M} \parallel {{\cal A}_{\lnot \Phi}}} = \obs X_{\cal M}$, and sequential update relation ${\rm Update}_{{\cal M} \parallel {{\cal A}_{\lnot \Phi}}}(s,q,r',q') = {\rm Update}_{\cal M}(s,r') \land {\rm Update}_{{\cal A}_{\lnot \Phi}}(\obs s, q, q')$. We extend the fair states of ${\cal A}_{\lnot \Phi}$ to ${\cal M} \parallel {{\cal A}_{\lnot \Phi}}$, i.e., we define them as $\{ (s, q) \mid s \in S, q \in F\}$, and as a result we have that $L^{\rm f}_{{\cal M} \parallel {{\cal A}_{\lnot \Phi}}} = L_{\cal M} \cap L^{\rm f}_{{\cal A}_{\lnot \Phi}} = L_{\cal M} \cap L_{\lnot \Phi}$. This reduces our language emptiness question to the equivalent {\em fair emptiness} problem $L^{\rm f}_{{\cal M} \parallel {\cal A}_{\lnot \Phi}} = \emptyset$.

The fair emptiness problem amounts to showing that all executions of ${\cal M} \parallel {\cal A}_{\lnot \Phi}$ are unfair, and we do so by presenting a ranking function that witnesses fair termination~\cite{DBLP:conf/icalp/LehmannPS81,DBLP:journals/iandc/GrumbergFR85}. 
A ranking function for fair termination is a map $V \colon \reg S \times Q \to R$ where $(R, \prec)$ defines a well-founded relation and, for all system and automaton states $\bs,\bs' \in S, \bq,\bq' \in Q$, the following two conditions hold true:
\begin{align}
    &(\bs,\bq) \rightarrow_{{\cal M}\parallel {\cal A}_{\lnot \Phi}} (\bs', \bq') \implies V(\reg \bs,\bq) \succeq V(\reg \bs',\bq')\label{eq:fairnoninc}\allowdisplaybreaks\\
    &(\bs,\bq) \rightarrow_{{\cal M}\parallel {\cal A}_{\lnot \Phi}} (\bs', \bq') \land \bq \in F \implies V(\reg \bs,\bq) \succ V(\reg \bs',\bq')\label{eq:fairdec}
\end{align}
A ranking function $V$ strictly decreases every time a transition from a fair state is taken, and never increases in any other case. Since every strictly decreasing sequence must be bounded from below (well-foundedness), every fair state can be visited at most finitely many times; the intuition is presented in Figure~\ref{fig:block-diagram}b, where \({\bf 1}_F(q)\) denotes the indicator function of $F$, returning 1 if \(q \in F\) and 0 otherwise. The existence of a valid ranking function represented in some form establishes that every execution of ${\cal M} \parallel {\cal A}_{\lnot \Phi}$ is necessarily unfair~\cite{DBLP:journals/apal/Vardi91}. In this work, we represent ranking functions as neural networks, the parameters of which we train from generated sample executions.

\begin{figure}
    \centering
    \begin{tikzpicture}
\node[draw,thick,minimum width=15mm, minimum height=7mm] (L) {Learn};
\node[draw,thick,minimum width=15mm, minimum height=7mm, right=3.5cm of L] (V) {Check};

\def\mdist{.7}

\node (M) at ($(L)!0.5!(V)+(0,\mdist)$) {${\cal M} \parallel {\cal A}_{\lnot \Phi}$};
\path (M) edge[-] node[fill=white, minimum width=0,minimum height=0] {$D$} ($(L)+(0,\mdist)$) ($(L)+(0,\mdist)$) edge[-stealth] (L);
\path (M) edge[-] ($(V)+(0,\mdist)$) ($(V)+(0,\mdist)$) edge[-stealth] (V);
\path (L) edge[-stealth] node[fill=white]{$\bar{V}, \theta$} (V);

\path (V) edge[-] ($(V)-(0,\mdist)$)
    ($(V)-(0,\mdist)$) edge[-] node[fill=white]{$D \leftarrow D \cup \{ {\tt cex} \} $} ($(L)-(0,\mdist)$)
    ($(L)-(0,\mdist)$) edge[-stealth] (L);

\path (V) edge[-stealth] node[fill=white] {$\tilde{V}, \tilde{\theta}$} ($(V.east)+(16mm,0)$) ($(V.east)+(16mm,0)$) node[right,anchor=west] {\tt valid!};
\end{tikzpicture}
    \caption{Learn-and-check workflow for \emph{provably sound} neural ranking function learning}
    \label{fig:traincheck}
\end{figure}
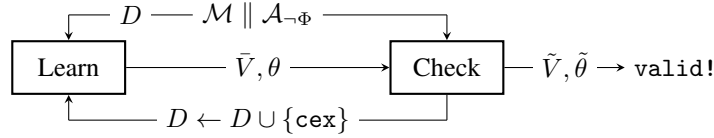

\section{Neural Ranking Functions for Fair Termination}
\label{sec:neuralnetwork}

We approach the problem of computing a ranking function for fair termination by training a neural network \(\bar{V} \colon \mathbb{R}^n \times \Theta \to \mathbb{R}\), with $n$ input neurons where \(n = |\reg X_{\cal M}|\) is the number of register variables of the system, 
one output neuron, and with a space of learnable parameters $\Theta$ for its weights and biases. We associate a distinct trainable parameter $\theta_q \in \Theta$ to each state $q \in Q$ of the B\"uchi automaton. We train these parameters on sampled executions of \({\cal M} \parallel {\cal A}_{\lnot \Phi}\) to ultimately represent 
a ranking function as a neural network $V(r, q) \equiv \bar{V}(r; \theta_q)$, which we call a neural ranking function. This scheme is illustrated in Figure~\ref{fig:block-diagram}, where we denote the set of all parameters by the unindexed~$\theta$.

We define our training objective as fulfilling conditions \eqref{eq:fairnoninc} and \eqref{eq:fairdec} on our synthetic dataset of sampled executions which, by analogy with reinforcement learning, can be viewed as a special case of episodes~\cite{DBLP:journals/jair/IcarteKVM22,DBLP:journals/ai/HasanbeigKA23}. Subsequently, we verify the conditions symbolically over the full state space \(S \times Q\) using satisfiability solving modulo theories (SMT)~\cite{DBLP:series/txtcs/KroeningS16,DBLP:series/faia/BarrettSST09}, to confirm the validity of our neural ranking function or obtain a counterexample for re-training. Overall, our approach combines learning and SMT-based checking for both efficacy and formal soundness, as illustrated in Figure~\ref{fig:traincheck}.

For a system $\cal M$ and a specification $\Phi$, we train the parameters $\theta$ of a neural network $\bar{V}$ 
from a sample dataset 
$D \subset \reg S \times Q \times \reg S \times Q$ of subsequent transition pairs, which we construct from random 
executions of the synchronous composition \({\cal M} \parallel {\cal A}_{\lnot \Phi}\). Each execution $(\bar{\bs}_0,\bar{\bq}_0), (\bar{\bs}_1,\bar{\bq}_1), \dots, (\bar{\bs}_k,\bar{\bq}_k)$ initiates from a random system and automaton state pair and is then simulated over a finite number of steps; the inputs to $\cal M$ and the non-deterministic choices in \({\cal A}_{\lnot \Phi}\) are resolved randomly. Our dataset~$D$ is constructed as the set of all quadruples \((\reg \bar{\bs}_i, \bar{\bq}_i, \reg \bar{\bs}_{i+1}, \bar{\bq}_{i+1})\) for $i = 0, \dots, k-1$ from all sampled executions, capturing consecutive state pairs along each execution; notably, the order in which quadruples are stored in $D$ is immaterial for our purpose, as our method reasons and trains locally on each transition pair regardless of their order of appearance along any execution.

We train the parameters of our neural network $\bar{V}$ to satisfy the ranking function conditions \eqref{eq:fairnoninc} and \eqref{eq:fairdec} over $D$. For each quadruple \((\br, \bq, \br', \bq') \in D\), 
this amounts to minimising the following loss function:
\begin{equation}
    {\cal L}_{\text{Rank}}(\br,\bq,\br',\bq'; \theta ) = \relu(\bar{V}(\br'; \theta_{\bq'}) - \bar{V}(\br; \theta_{\bq}) + \epsilon \cdot {\bf 1}_F(\bq)).
\end{equation}
where $\epsilon > 0$ is a hyper-parameter that denotes the margin for the decrease condition. When ${\cal L}_{\text{Rank}}$ takes its minimum value---which is zero---then the following two cases are satisfied:
if \(\bq \not\in F\), then $\bar{V}$ does not increase along the given transition, i.e., \(\bar{V}(\br; \theta_{\bq}) \geq \bar{V}(\br'; \theta_{\bq'})\), which corresponds to satisfy condition \eqref{eq:fairnoninc}; 
if otherwise \(\bq \in F\), then $\bar{V}$ decreases by at least the margin $\epsilon > 0$ along the given transition, i.e., \(\bar{V}(\br; \theta_{\bq}) \geq \bar{V}(\br'; \theta_{\bq'}) + \epsilon\),  which corresponds to satisfy condition \eqref{eq:fairdec}.

Overall, our learning phase ensures that the total loss function \({\cal L}(D; \theta)\) below takes value zero:
\begin{equation}
    \mathcal{L}(D; \theta) = \mathbb{E}_{(\br,\bq,\br',\bq') \in D}[{\cal L}_{\text{Rank}}(\br,\bq,\br',\bq'; \theta) ]
\end{equation}
Unlike many other machine learning applications, for our purpose it is essential to attain the global minimum; if this fails, there are counterexamples to $\bar V$ being a ranking function in the dataset \(D\) itself. To facilitate the optimisation process, we train the parameters associated to 
each automaton state independently, one after the other, as opposed to training all parameters at once. Iteratively, we select one automaton state $q \in Q$ and optimise only $\theta_q \in \Theta$ for a number of steps, while keeping all other parameters $\theta_{q'} \in \Theta$ fixed to their current value, for all $q' \neq q$. We repeat the process over each automaton state, possibly iterating over the entire set of automaton states $Q$ multiple times, until the total loss $\mathcal{L}(D; \theta)$ takes value zero. 

\begin{figure}
\centering
\begin{tikzpicture}[minimum size=11pt,node distance=5mm,inner sep=0pt]

    \def\vdist{5mm}
    \def\hdist{15mm}
    \def\circWidth{0.35mm}

    \foreach \x in {1,2,4}{
        \node[draw, circle, line width=\circWidth] (first\x) at (0,\vdist*-\x+\vdist) {};
    }
    \node[] at ($(first2)!0.4!(first4)$) {\vdots};

    \foreach \x in {1,2,4}{
        \node[draw, circle, line width=\circWidth] (second\x) at (\hdist,\vdist*-\x+\vdist) {};
    }
    \node[] at ($(second2)!0.4!(second4)$) {\vdots};

    \foreach \x in {1,2,4}{
        \node[draw, circle, line width=\circWidth] (third\x) at (2*\hdist,\vdist*-\x+\vdist) {};
    }
    \node[] at ($(third2)!0.4!(third4)$) {\vdots};

    \def\layersep{3.5mm}

    % Fully Connected Layer 1
    \foreach \x in {1,2,4}{
        \node[draw, circle, line width=\circWidth] (firsttrain\x) at (3*\hdist,\vdist*-\x+\vdist) {};
    }
    \node[] at ($(firsttrain2)!0.4!(firsttrain4)$) {\vdots};
    \begin{scope}[rounded corners=6pt]
        \draw[draw] ($(firsttrain1)+(-\layersep,+\layersep)$) rectangle ($(firsttrain4)+(+\layersep,-\layersep)$);
    \end{scope}
    
    % Fully Connected Layer 2
    \foreach \x in {1,2,4}{
        \node[draw, circle, line width=\circWidth] (secondtrain\x) at (4*\hdist,\vdist*-\x+\vdist) {};
    }
    \node[] at ($(secondtrain2)!0.4!(secondtrain4)$) {\vdots};
    \begin{scope}[rounded corners=6pt]
        \draw[draw] ($(secondtrain1)+(-\layersep,+\layersep)$) rectangle ($(secondtrain4)+(+\layersep,-\layersep)$);
    \end{scope}
    
    % Intermediate Layers
    \node at (5*\hdist, \vdist*-3+\vdist+0.5*\vdist) {\dots};

    % Last Fully Connected Layer
    
    \foreach \x in {1,2,4}{
        \node[draw, circle, line width=\circWidth] (lasttrain\x) at (6*\hdist,\vdist*-\x+\vdist) {};
    }
    \node[] at ($(lasttrain2)!0.4!(lasttrain4)$) {\vdots};
    \begin{scope}[rounded corners=6pt]
        \draw[draw] ($(lasttrain1)+(-\layersep,+\layersep)$) rectangle ($(lasttrain4)+(+\layersep,-\layersep)$);
    \end{scope}
    % Output Layer

    \node[draw,circle, line width=\circWidth] (output) at (7*\hdist,-3*\vdist+\vdist+0.5*\vdist) {};

    % Connections between layers
    \foreach \x in {1,2,4}{
        \draw [thick] (first\x) edge[-] (second\x);
        \draw [thick] (second\x) edge[-] (third\x);
        \foreach \y in {1,2,4} {
            \draw [thick] (third\x) edge[-] (firsttrain\y);
        }
    }

    \foreach \x in {1,2,4}{
        \foreach \y in {1,2,4} {
            \draw [thick] (firsttrain\x) edge[-] (secondtrain\y);
        }
        \draw [thick] (lasttrain\x) edge[-] (output);
    }

 \node[below right=0.2cm and 0.2cm of first4] {\footnotesize Norm.};
cm of third1] {\footnotesize   NoAct.};
 \node[above =0.2cm of firsttrain1] {\footnotesize  
 $\clamp$};
 \node[above =0.2cm of secondtrain1] {\footnotesize  $\clamp$};
\node[above =2mm of lasttrain1] {\footnotesize  $\clamp$};

\def\brdist{3mm}%

\draw[decorate,decoration={brace,amplitude=6pt,mirror,raise=0pt}] let
    \p1=(second4.east),\p2=(output.west),\p3=(secondtrain4.south) in (\x1,\y3-\brdist) --
    (\x2,\y3-\brdist) node[midway,yshift=-5mm]{Trainable parameters $\theta_q$};

\draw[decorate,decoration={brace,amplitude=6pt,mirror,raise=0pt}] let
    \p1=(first1.north),\p2=(first4.south) in (\x1-\brdist,\y1) --
    (\x1-\brdist,\y2) node[midway,left,xshift=-2mm]{r};

\node[right=1mm of output] {$\bar{V}$};
\end{tikzpicture}
\caption{Neural ranking function architecture} 
\label{fig:arch}
\end{figure}
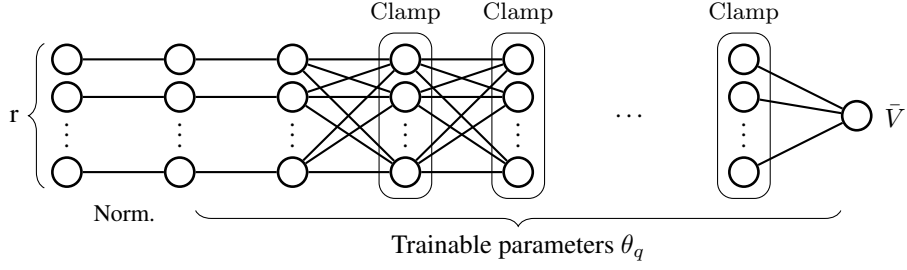

Our neural network $\bar V$ follows a feed-forward architecture as depicted
in Figure~\ref{fig:arch}: for a given automaton state $q \in Q$ and
associated parameter $\theta_q$, it takes an $n$-dimensional input $r \in
\mathbb{R}^n$ where each input neuron corresponds to the value of a register
variable in $\reg X_{\cal M}$, and produces one output for the corresponding
ranking value $\bar{V}(r; \theta_q)$.  Our architecture consists of a
normalisation layer, followed by an element-wise multiplication layer, in
turn followed by a multi-layer perceptron with clamped ReLU activation
functions.  The first layer applies a scaling factor to each input neuron
independently to ensure consistent value ranges across inputs, implemented
via element-wise multiplication with a constant vector of scaling
coefficients derived from the dataset $D$ before training; this integrates
data normalisation into the network, enables $\bar{V}$ to use raw data from
\({\cal M}\) and simplifies the symbolic encoding of the normalisation
operation during the verification phase.  The second layer applies a
trainable scaling factor to each individual neuron and is implemented via
element-wise multiplication with a $n$-dimensional vector with trainable
coefficients.  Finally, this is followed by a fully connected multi-layer
perceptron with trainable weights and biases, with the activation function
defined as the element-wise application of $\clamp(x; u) = \max(0, \min(x,
u))$; the upper bound $u$ and the depth and width of the hidden layers of
the multi-layer perceptron component are hyper-parameters chosen to optimise
training and verification performance.

Attaining zero total loss ${\cal L}(D;\theta)$ guarantees that our neural
ranking function candidate $\bar{V}$ satisfies the ranking criteria for fair
termination over the dataset $D$ but not necessarily over the entire
transition relation $\rightarrow_{{\cal M}\parallel {\cal A}_{\lnot \Phi}}$,
as required to fulfil conditions \eqref{eq:fairnoninc} and
\eqref{eq:fairdec} and consequently to answer our model checking question
(cf.~Section~\ref{sec:Automata}).  To formally check whether the ranking
criteria are satisfied over the entire transition relation, we couple our
learning procedure with a sound decision procedure that verifies their
validity, as illustrated in Figure~\ref{fig:traincheck}.

We check the validity of our candidate ranking neural network using
satisfiability modulo the theory of bit-vectors.  While the sequential
update relation $\text{Update}_{{\cal M} \parallel {\cal A}_{\lnot \Phi}}$
is natively expressed over the theory of bit-vectors, the formal semantics
of the neural network $\bar{V}$ is defined on the reals.  Hence, encoding
$\bar{V}$ and $\text{Update}_{{\cal M} \parallel {\cal A}_{\lnot \Phi}}$
within the same query would result in a combination of real and bit-vector
theories, which is supported in modern SMT solvers but often leads to
sub-optimal performance~\cite{DBLP:series/txtcs/KroeningS16}.  Therefore, to
leverage the efficacy of specialised solvers for the theory of
bit-vectors~\cite{DBLP:conf/cav/NiemetzP23}, we quantise our neural network
using a standard approach for this
purpose~\cite{DBLP:conf/cvpr/JacobKCZTHAK18}; this converts all arithmetic
operations within the neural networks into fixed-point arithmetic, which are
implemented using integer arithmetic only.  We quantise our parameters to
their respective integer representation $\tilde{\theta} \approx 2^{\fr}
\cdot \theta$, where $\fr$ is a hyper-parameter for the number of fractional
digits in fixed-point representation, and we replace linear layers and
activation functions by their quantised counterpart; readers may consult the
relevant literature for more detailed information on neural network
quantisation~\cite{DBLP:conf/cvpr/JacobKCZTHAK18,DBLP:conf/tacas/GiacobbeHL20}. 
This results in a quantised neural network $\tilde{V} \colon \mathbb{Z}^n
\times \tilde{\Theta} \to \mathbb{Z}$ that approximates our trained network
$\tilde{V} \approx 2^{\fr} \cdot \bar{V}$, where $\tilde{\Theta}$ denotes
the space of integer parameters.  
fractional digits introduced by the linear
layers~\cite{DBLP:conf/cvpr/JacobKCZTHAK18,DBLP:conf/tacas/GiacobbeHL20}. 
We consider the quantised network $\tilde{V}$ as our candidate proof
certificate for fair termination.

We reduce the validity query---whether our quantised neural network
$\tilde{V}$ satisfies the ranking criteria for fair termination
\eqref{eq:fairnoninc} and \eqref{eq:fairdec} over the entire transition
relation of ${\cal M} \parallel {\cal A}_{\lnot \Phi}$---to the dual
satisfiability query for the existence of a counterexample to the criteria. 
Specifically, we delegate to an off-the-shelf SMT solver the task of
computing a satisfying assignment $s \in S, r' \in \reg S$ for which the
following quantifier-free first-order logic formula is satisfied:
\begin{equation}
   \bigvee_{q,q'\in Q} 
   {\rm Update}_{{\cal M} \parallel {\cal A}_{\lnot \Phi}}(s, q, r', q') \land 
    \tilde{V}(\reg s; \tilde{\boldsymbol{\theta}}_q) - {\bf 1}_{F}(q) <  \tilde{V}(r'; \tilde{\boldsymbol{\theta}}_{q'}) \label{eq:verif}
\end{equation}
where $\tilde{\boldsymbol{\theta}}$ is the (constant) parameter resulting from training and quantisation. 
We encode the quantised neural network $\tilde{V}$ using a standard translation into first-order logic over the theory of bit-vectors~\cite{DBLP:conf/tacas/GiacobbeHL20}, supplementing it with specialised rewriting rules to enhance the solver's performance, as detailed in Appendix~\ref{sec:apd_quantNN}. We additionally note that $\tilde{V}$ is guaranteed to be bounded from below as $S$ is finite, albeit potentially very large, i.e., exponential in the combined bit-width of $X_{\cal M}$.

If the solver finds a satisfying assignment, then the assignment represents a transition of $\cal M$ that 
refutes the validity of $\tilde{V}$; in this case, we extend it to a respective transition in ${\cal M} \parallel {\cal A}_{\lnot \Phi}$, we add it to our dataset $D$ and repeat training and verification in a loop. 
Conversely, if the solver determines that formula \eqref{eq:verif} is unsatisfiable, then our procedure 
concludes that $\tilde{V}$ is formally a valid neural ranking function and, consequently, system $\cal M$ satisfies specification $\Phi$. 

We note that LTL model checking of hardware designs is decidable and PSPACE-complete~\cite{mc2nd,principlesofmc,principlesofcps}. While it is theoretically possible for our approach to achieve completeness when a ranking function exists by enumerating all transitions and employing a sufficiently large neural network as a lookup table over the entire state space, this is impractical for all but toy cases. 
In this work, we employ tiny neural networks and incomplete but practically effective gradient 
descent algorithms to train neural ranking functions. We experimentally demonstrate on a large set of formal hardware verification benchmarks that this solution is very effective in practice. 

\section{Illustrative Example}
\label{sec:case_studies}

\begin{figure}
    \centering
    \begin{tabular}{cc}
        \begin{minipage}{0.42\textwidth}
            \centering
            \input{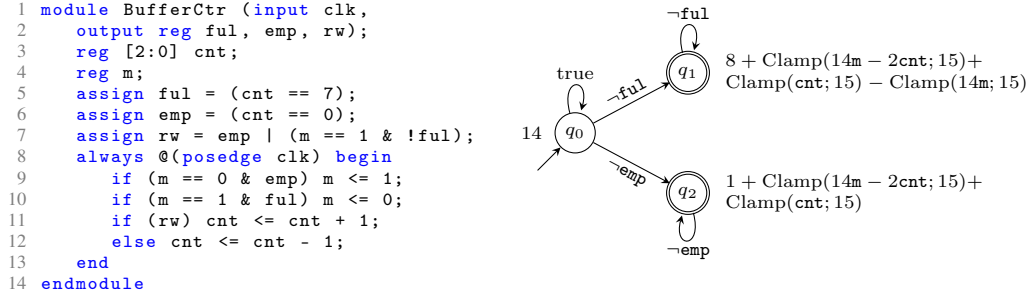}
        \end{minipage} &
        \begin{minipage}{0.47\textwidth}
            \centering
            \begin{tikzpicture}[>=stealth,on grid,initial text={}]
    \scriptsize
    \node[state, minimum size=0.5cm] (nAcc) {$q_0$};
    \node[state, double, minimum size=0.5cm] (Acc1) [above right= 0.8cm and 1.5cm of nAcc] {$q_1$};
    \node[state, double, minimum size=0.5cm] (Acc2) [below right= 0.8cm and 1.5cm of nAcc] {$q_2$};

     % Short arrow from a point left nAcc
    \path[->] ($(nAcc)-(0.5,0.5)$) edge (nAcc);

     % Adding the equation above Acc1
   \node[right= 2mm of Acc1,anchor = west] (eq1) { \begin{tabular}{l}
        $8 + \clamp(14{\tt m} - 2{\tt cnt}; 15) +$ \\ $ \clamp({\tt cnt}; 15) - \clamp(14{\tt m}; 15)$
   \end{tabular}};

    \node[right= 2mm of Acc2, anchor = west] (eq1) { \begin{tabular}{l}
       $1 + \clamp(14{\tt m} - 2{\tt cnt}; 15) + $ \\ $ \clamp({\tt cnt}; 15)$
   \end{tabular}};

    \node[left= 3.5mm of nAcc, anchor = east] (eq0) {$14$ };

    \path[->]
        (nAcc) edge [loop above] node [text width=1.5cm,align=center] { $\rm true$} ()
        (Acc1) edge [loop above] node [above] {$ \lnot {\tt ful}$} ()
        (Acc2) edge [loop below] node [below] { $ \lnot {\tt emp}$} ()
        (nAcc) edge node [text width=1.7cm,align=center, above, sloped] { $\lnot {\tt ful}$} (Acc1)
        (nAcc) edge node [text width=1.7cm,align=center,below, sloped] { $\lnot {\tt emp}$} (Acc2);
\end{tikzpicture}
        \end{minipage} \\
    \end{tabular}
    \caption{Illustrative hardware design, B\"uchi automaton, and respective ranking function}
    \label{fig:illustrativeEx}
\end{figure}
Modern hardware designs frequently incorporate word-level arithmetic operations, the simplest of which being counter increments/decrements, which are a staple in hardware engineering~\cite{DBLP:conf/hotos/ZhangDFS07, maini2007digital}. 
One such example is illustrated as part of the SystemVerilog module in Figure~\ref{fig:illustrativeEx}.
This represents a simplified buffer controller that counts the number of packets 
stored in the buffer and indicates when the buffer is full or empty with the  {\tt ful} and {\tt emp} signals, respectively. This specific controller internally coordinates read-and-write operations through the {\tt rw} signal: 
iteratively, the system signals {\tt rw} = 1 until the buffer is full and then {\tt rw} = 0 until the buffer is empty.

The design satisfies the property that both our
observables {\tt ful} and {\tt emp} are true infinitely
often, captured by the LTL formula
$\Phi = \ltlGF {\tt ful} \land \ltlGF {\tt emp}$.
Dually, this specification says that the system does not eventually go into a state from where $\lnot {\tt ful}$ holds
indefinitely nor $\lnot {\tt emp}$ holds indefinitely, that is,
$\lnot \Phi = \ltlFG \lnot {\tt ful} \lor \ltlFG \lnot {\tt emp}$.
Equivalently, this amounts to proving that no system trace is in the fair language of the automaton ${\cal A}_{\lnot \Phi}$ given in Figure~\ref{fig:illustrativeEx}. 

A neural ranking function $\bar{V}$ for the fair termination of this system and automaton has 
5 input neurons for the register variables {\tt cnt}, {\tt m}, {\tt ful}, {\tt emp}, 
and {\tt rw}, and one hidden layer with three neurons in the multi-layer perceptron component. As illustrated in Figure~\ref{fig:illustrativeEx}, 
each automaton state is associated with a ranking function defined in terms of this architecture and their respective parameters. The sequence below gives an execution of model states alongside the respective ranking function values:
\begin{table}[h]
    \centering
    \setlength{\tabcolsep}{2.7pt}
    \begin{tabular}{c|c|*{22}{c|}c}
      \multicolumn{1}{c}{} & \multicolumn{3}{c}{~~~~~~~\tt emp} & \multicolumn{4}{c}{} & \multicolumn{3}{c}{\tt ful} & \multicolumn{4}{c}{} & \multicolumn{3}{c}{\tt emp} & \multicolumn{4}{c}{} & \multicolumn{3}{c}{~~~~\tt ful}\\
\cline{2-24}
     &{\tt cnt} & 0 & 1 & 2 & 3 & 4 & 5 & 6 & 7 & 6 & 5 & 4 & 3 & 2 & 1 & 0 & 1 & 2 & 3 & 4 & 5 & 6 & 7 & \\\cline{2-24}
     &{\tt m} & 0 & 1 & 1 & 1 & 1 & 1 & 1 & 1 & 0 & 0 & 0 & 0 & 0 & 0 & 0 & 1 & 1 & 1 & 1 & 1 & 1 & 1\\\cline{2-24}
     &{\tt rw} & 1 & 1 & 1 & 1 & 1 & 1 & 1 & 0 & 0 & 0 & 0 & 0 & 0 & 0 & 1 & 1 & 1 & 1 & 1 & 1 & 1 & 0 \\\cline{2-24}
     &\scriptsize $\bar{V}(\cdot; \theta_{q_0})$ & 14 & 14 & 14 & 14 & 14 & 14 & 14 & 14 & 14 & 14 & 14 & 14 & 14 & 14 & 14 & 14 & 14 & 14 & 14 & 14 & 14 & 14 \\\cline{2-24}
     &\scriptsize $\bar{V}(\cdot; \theta_{q_1})$ & 8 & 7 & 6 & 5 & 4 & 3 & 2 & 1 & 14 & 13 & 12 & 11 & 10 & 9 & 8 & 7 & 6 & 5 & 4 & 3 & 2 & 1\\\cline{2-24}
     &\scriptsize $\bar{V}(\cdot; \theta_{q_2})$ & 1 & 14 & 13 & 12 & 11 & 10 & 9 & 8 & 7 & 6 & 5 & 4 & 3 & 2 & 1 & 14 & 13 & 12 & 11 & 10 & 9 & 8\\\cline{2-24}
\end{tabular}
\end{table}

One can observe that all transitions throughout this execution satisfy conditions \eqref{eq:fairnoninc}
 and \eqref{eq:fairdec}. This assessment is based on the (not explicitly presented) synchronous composition with the automaton. First, we note that every transition originating from $q_0$ has a non-increasing ranking value, as $\bar{V}(\cdot; \theta_{q_0}) = 14$ is an upper bound to all other values. Furthermore, every transition leaving $q_1$—that is, every transition whose source state satisfies $\lnot {\tt ful}$—exhibits a strictly decreasing value $\bar{V}(\cdot; \theta_{q_1})$. Similarly, the same observation applies to $q_2$ and the condition $\lnot {\tt emp}$. We note that the
 transitions that exhibit increasing values from 1 to 14 in this execution are impossible over the synchronous composition; this is because they are originating from states that satisfy both ${\tt ful}$ and $q_1$, and similarly states that satisfy both ${\tt emp}$ and $q_2$, and which do not have corresponding transitions in the automaton.

This neural ranking function admits no increasing transition originating from $q_0$ and no non-decreasing transitions originating from $q_1$ or $q_2$ on the synchronous composition of the system and the automaton. Therefore, it is a valid proof certificate for every system trace to satisfy specification $\Phi$.

\section{Experimental Evaluation}
\label{sec:result}

\begin{figure}
    \centering
    \begin{tabular}{ccc}\hspace{0.25cm}
         \begin{picture}(165,102.5)
            \put(0,0){\includegraphics[width=165pt, height=102.5pt]{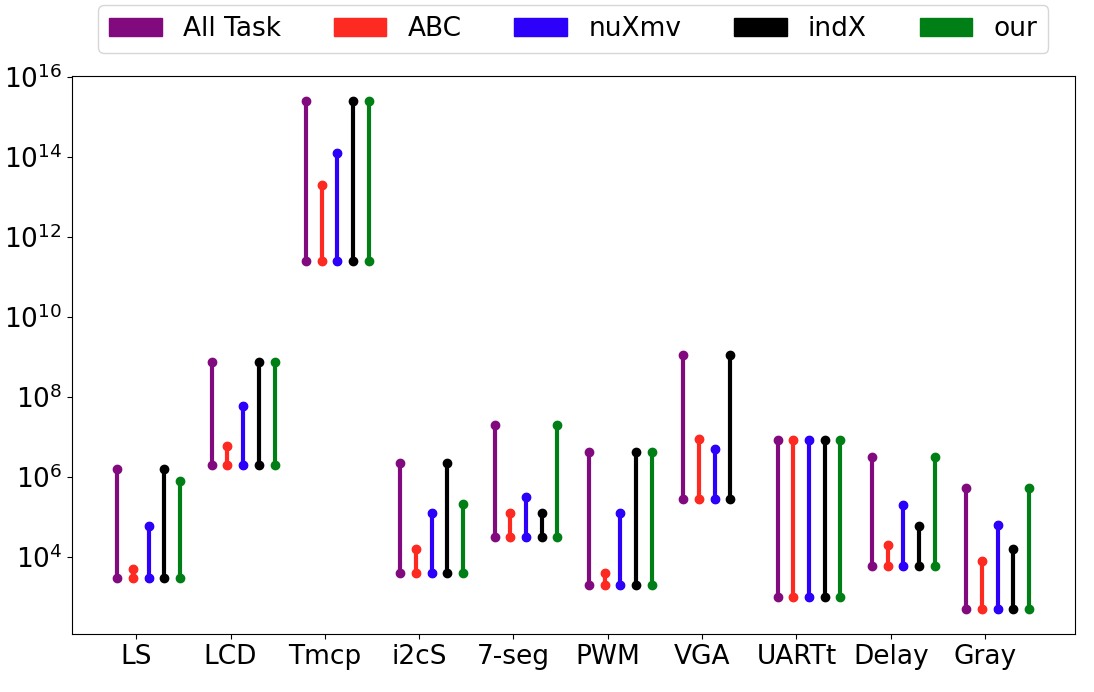}} 
            \put(-6,34){\rotatebox{90}{\tiny State Space Size}}
        \end{picture}& &
        \begin{picture}(165,102.5)
            \put(0,0){\includegraphics[width=165pt, height=102.5pt]{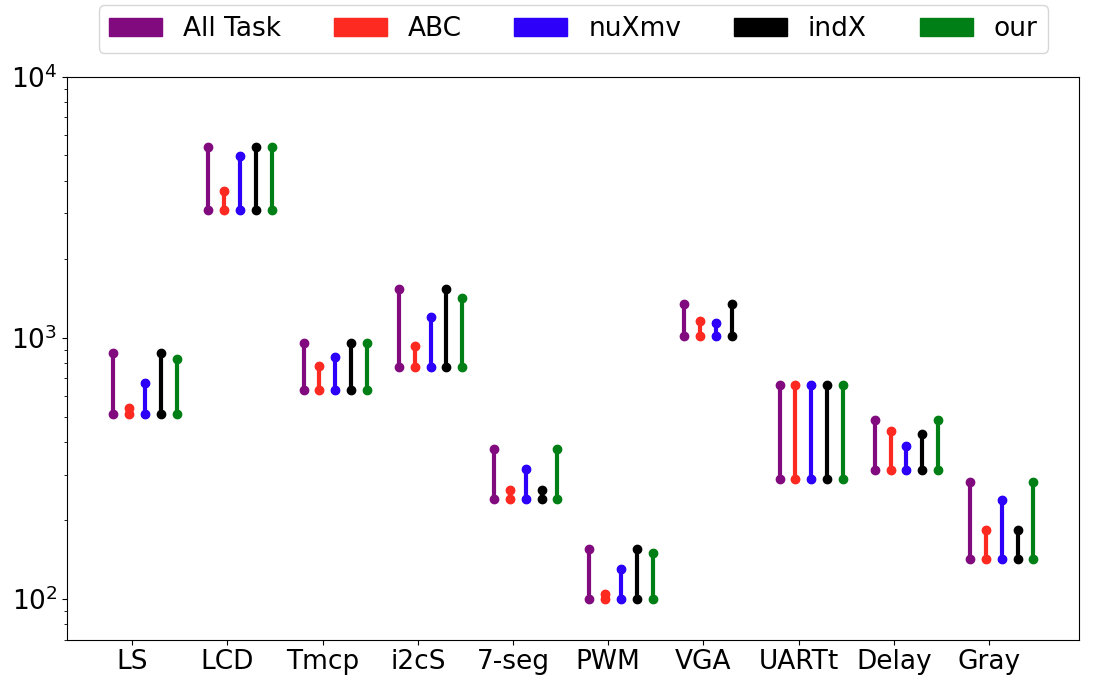}} 
            \put(-6,32){\rotatebox{90}{\tiny Logic Gate Count}}
        \end{picture}
    \end{tabular}
    \caption{ Solved tasks in terms of state space size and logic gate count (log scale)}
    \label{fig:cmp_case}
\end{figure}

We examine 194 verification tasks derived from ten parameterised hardware
designs, detailed in Appendix~\ref{sec:benchmarks}.  By adjusting parameter
values, we create tasks of varying complexity, resulting in different logic
gate counts and state space sizes, thus offering a broad spectrum of
verification complexity for tool comparison.  The parameter ranges for each
design are given as “all tasks” in Figure~\ref{fig:cmp_case}.  These tasks
serve as benchmarks to evaluate the scalability of our method relative to
conventional model checking.

\paragraph{Implementation} We have developed a prototype tool for neural
model checking\footnote{\url{https://github.com/aiverification/neuralmc}},
utilising Spot 2.11.6~\cite{DBLP:conf/cav/Duret-LutzRCRAS22} to generate the
automaton $\mathcal{A}_{\lnot \Phi}$ from an LTL specification $\Phi$.  As
depicted in Figure~\ref{fig:block-diagram}, the circuit model $\mathcal{M}$
and the automaton $\mathcal{A}_{\lnot \Phi}$ synchronise over a shared clock
to form a product machine.  Using Verilator version
5.022~\cite{snyder2004verilator}, we generate a dataset $D$ from finite
trajectories of this machine.  This dataset trains a neural network using
PyTorch  2.2.2, as outlined in Section~\ref{sec:neuralnetwork}.  To ensure
formal guarantees, the network is quantised and subsequently translated to
SMT, following the process outlined in Appendix~\ref{sec:apd_quantNN}.  The
SystemVerilog model is converted to SMT using EBMC
5.2~\cite{DBLP:conf/isvlsi/MukherjeeKM15}.  We check the satisfiability
problem using the Bitwuzla 0.6.0 SMT solver~\cite{DBLP:conf/cav/NiemetzP23}.

\paragraph{State of the Art} We benchmarked our neural model checking
approach against two leading model checkers,
nuXmv~\cite{DBLP:conf/cav/CavadaCDGMMMRT14} and
ABC~\cite{DBLP:conf/cav/BraytonM10, superProve}.  ABC and nuXmv were the top
performers in the liveness category of the hardware model checking
competition (HWMCC)~\cite{biere2017hardware,bierehardware}.
Our comparison employed
the latest versions: nuXmv 2.0.0 and ABC's Super Prove tool
suite~\cite{superProve}, which were also used in the most recent HWMCC'20~\cite{bierehardware}.  We
further consider two widely used industrial formal verification tools for
SystemVerilog, anonymised as industry tool~X and industry tool~Y.  Tool Y
fails to complete any of the 194 tasks and is therefore not referenced
further in this section.

\paragraph{Experimental Setup}  Evaluations were conducted on an Intel Xeon
2.5\,GHz processor with eight threads and 32\,GB of RAM running Ubuntu 20.04. 
Bitwuzla and nuXmv utilise one core each, ABC used three cores, and PyTorch
leveraged all available cores.  Each tool was allotted a maximum of five hours
for each verification task, as detailed in Appendix~\ref{sec:all_exp}.

\paragraph{Hyper-parameters} We instantiate the architecture described in
Section~\ref{sec:neuralnetwork} and illustrated in Figure~\ref{fig:arch},
employing two hidden layers containing 8 and 5 neurons.  The normalisation
layer scales the input values to the range [0, 100].  We train with the
AdamW optimiser~\cite{DBLP:conf/iclr/LoshchilovH19}, typically setting the
learning rate to 0.1 or selecting from {0.2, 0.05, 0.03, 0.01} if adjusted,
with a fixed weight decay of 0.01, demonstrating minimal hyperparameter
tuning for training.

\paragraph{Dataset Generation} In hardware design, engineers utilise test
benches to verify safety properties through directed testing or Constraint
Random Verification (CRV), aiming for high coverage and capturing edge
cases~\cite{wilson, spearsystemverilog}.  We apply CRV to the SystemVerilog
file, generating random trajectories.  As outlined in Section
\ref{sec:neuralnetwork}, we start these trajectories by selecting the
internal states of model $\mathcal{M}$ (e.g., {\tt module BufferCtr} and
automaton $A_{\lnot\Phi}$; in Figure~\ref{fig:illustrativeEx}) using a
uniform distribution.  At each step, we assign random inputs to model
$\mathcal{M}$ and handle the non-determinism in automaton $A_{\lnot\Phi}$ by
making choices from uniform or skewed distributions.  We skew the
distribution when a particular event is too predominant or too rare.  In our
experiments, such skewing is rare and limited to the reset and enable
signals in $\mathcal{M}$, as well as the non-determinism in the automaton
$A_{\lnot\Phi}$.

\newcommand{\gt}[1]{\textcolor{gray}{#1}}
\newcommand{\Bf}[1]{\textbf{#1}}
\begin{table}
\centering % Center the table
\caption{Number of verification task completed by academic and industrial tool, per design}
\small
\label{tab:case_study}
\begin{tabular}{|l|*{11}{c|}} % Creates a table with 11 centered columns
\hline
      & LS      & LCD     & Tmcp    & i2cS    & 7-Seg   & PWM     & VGA     & UARTt   & Delay   &  Gray   & Total  \\ \hline
Tasks & 16      & 14      &   17    &   20    &   30    &   12    &   10    &   10    &   32    &   33    & 194    \\ \hline\hline
ABC   & \gt{2}  & \gt{3}  & \gt{7}  & \gt{3}  & \gt{8}  & \gt{2}  & \gt{3}  & \Bf{10} & \gt{6}  & \gt{13}  & 57     \\ \hline
nuXmv & \gt{8}  & \gt{9}  & \gt{12} & \gt{10} & \gt{10} & \gt{7}  & \gt{3}  & \Bf{10} & \gt{24} & \gt{24}  & 117     \\ \hline
our   & \gt{15} & \Bf{14} & \Bf{17} & \Bf{18} & \Bf{30} & \gt{11} & \gt{0}  & \Bf{10} & \Bf{32} & \Bf{33} &  180   \\ \hline\hline
Ind. X& \Bf{16} & \Bf{14} & \Bf{17} & \Bf{18} & \gt{18} & \Bf{12} & \Bf{10} & \Bf{10} & \gt{19} & \gt{22}  & 156   \\ \hline
Ind. Y& \gt{0}  & \gt{0}  &  \gt{0} & \gt{0}  & \gt{0}  & \gt{0}  & \gt{0}  & \gt{0}  & \gt{0}  & \gt{0}  & 0  \\ \hline

\end{tabular}
\end{table}

\paragraph{Solved Tasks} Table~\ref{tab:case_study} presents the number of completed tasks for each tool across the ten hardware designs, while Figure~\ref{fig:cmp_case} shows the range of state-space sizes and logic gate counts each tool successfully handled. Overall, our tool performs favourably in comparison to others, with the notable exception of the {\tt VGA} design, where training a ranking function failed due to local minima, preventing convergence to zero loss—a known limitation of gradient descent-based methods.

\paragraph{Aggregate Runtime Comparison}  

\begin{figure}
    \centering
    \begin{tabular}{ccc}
    \begin{picture}(128,85)
            \put(0,0){\includegraphics[width=128pt, height=92.5pt]{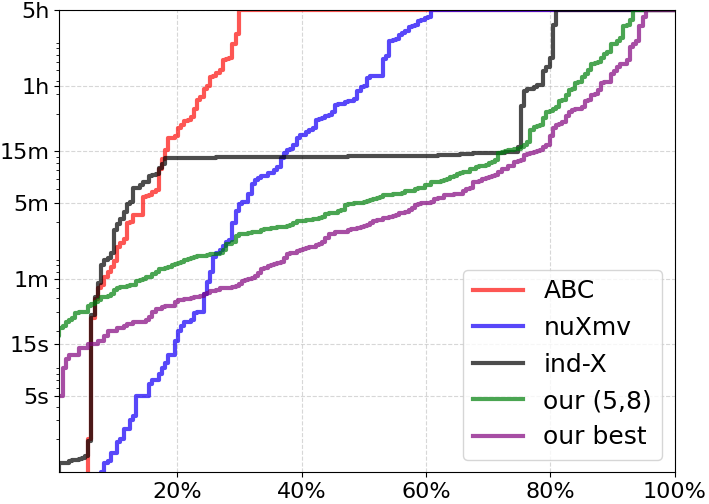}} 
            \put(50,-6){\tiny Tasks completed}
            \put(-3,43){\rotatebox{90}{\tiny Time}}
        \end{picture}
         & 
         \begin{picture}(125,85)
            \put(-5,0){\includegraphics[width=125pt, height=90pt]{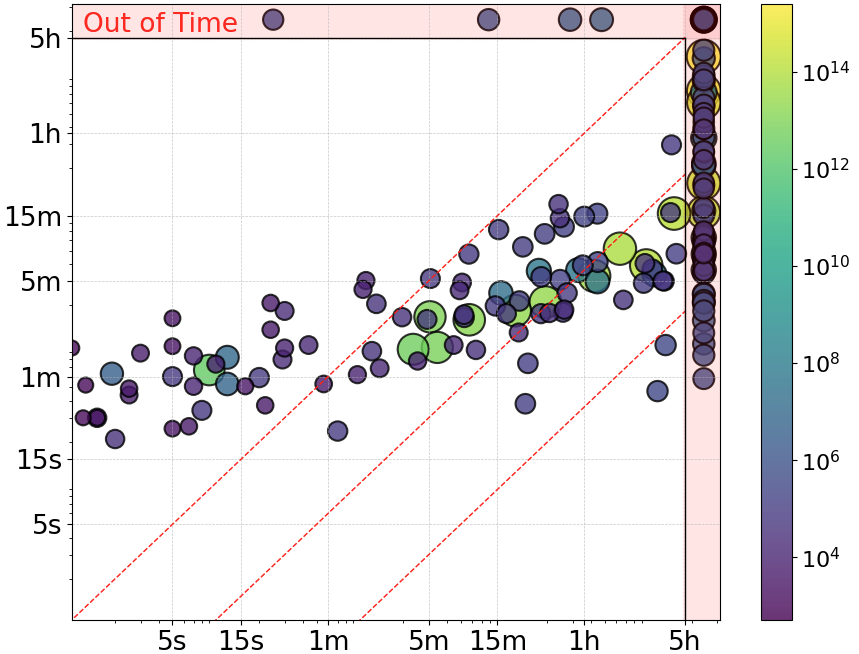}} 
            \setlength{\fboxsep}{0.5pt}
            \put(24,-6){\tiny Best of nuXmv/ABC time}
            \put(-10,38){\rotatebox{90}{\tiny Our time}}
            \put(9,10){\tiny \colorbox{white}{1X}}
            \put(29,10){\tiny \colorbox{white}{10X}}
             \put(49.5,10){\tiny \colorbox{white}{100X}}
            \put(120,29){\rotatebox{90}{\tiny State space size}}
        \end{picture}
        &
         \begin{picture}(115,85)
            \put(0,0){\includegraphics[width=115pt, height=92.5pt]{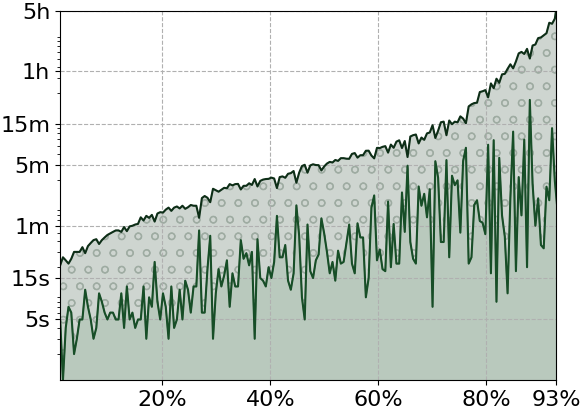}} 
            \put(43.5,-6){\tiny Tasks completed}
            \put(46,46){\tiny Total time}
            \put(42,10){\tiny Learning time}
           
            \put(-5,40){\rotatebox{90}{\tiny Our time}}
        \end{picture}\\\\
        (a) & (b) & (c)
    \end{tabular}
    \caption{Runtime comparison with the state of the art (all times are in log scale)}
    \vspace{-5mm}
    \label{fig:experiments}
\end{figure}
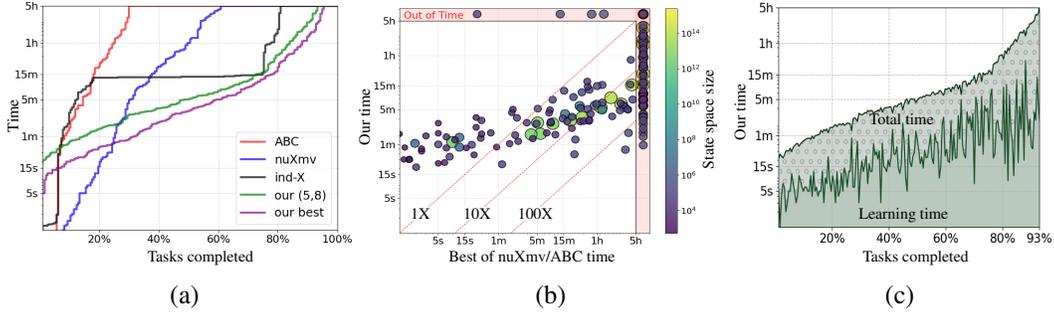

Figure~\ref{fig:experiments}a displays a cactus plot with a \qty{5}{\hour}
limit, we consider our configuration with 8 and 5 hidden neurons as detailed
in the section, along with the aggregate of the best time on individual
tasks obtained from our ablation study, as detailed in
Appendix~\ref{sec:apd_ablation_study}.  While the default
architecture performs the best across all tasks, on some tasks a smaller
network is sufficient, and leads to lower verification time.  At the same
time, larger networks often succeed on tasks that otherwise fail, making the
``our best'' line strictly better than ``our (5, 8)''.  This shows that
improvement can be obtained by tuning the width of the hidden layers; note that this analysis considers three additional configurations (i.e, (3, 2), (5, 3), (15, 8)) that adhere to the architecture introduced in Section~\ref{sec:neuralnetwork}.  For the rest of our experiments, we continue using the default architecture. 
The plot further shows that our tool completes \qty{93}{\percent} of tasks,
outperforming ABC, nuXmv, and industry tool~X, which completes
\qty{29}{\percent}, \qty{60}{\percent}, and \qty{80}{\percent},
respectively.  At any point in the time axis, we compute the difference
between the percentage of tasks completed by our tool with each of the
others in the figure.  Then, taking the average of these differences across
the time axis, showing that our method is successful in \qty{60}{\percent}
more tasks than ABC, \qty{34}{\percent} more than nuXmv, and
\qty{11}{\percent} more than the leading commercial model checker at any
given time.  Furthermore, the number of tasks completed by nuXmv in
\qty{5}{\hour} are finished by our tool in less than \qty{8}{\minute}, and
those completed by ABC in \qty{5}{\hour} take just under \qty{3}{\minute}
with our method.

\paragraph{Individual Runtime Comparison} Figure~\ref{fig:experiments}b presents a scatter plot where each point represents a verification task, with size and brightness
indicating the state-space size.  Points are plotted horizontally by the
lesser of time taken by nuXmv or ABC and vertically by our method's time. 
The plot reveals that academic tools time out on \qty{39}{\percent} of
tasks, while our method times out on \qty{7}{\percent}.  Moreover, we are
faster than the academic tools on \qty{67}{\percent} of tasks, 10 times
faster on \qty{34}{\percent}, and 100 times faster on \qty{4}{\percent}. 
These results demonstrate that we generally outperform the state of the art
on this benchmark set (see Appendix~\ref{tab:all_exp} for individual
runtimes).  However, we perform relatively worse on the {\tt UARTt} design. 
This design involves an $N$-bit register for data storage and a counter for
transmitted bits, enabling sequential outputs.  Since there is no word-level
arithmetic over the $N$-bit register, increasing its size minimally affects
the complexity of symbolic model checking.  Consequently, ABC, nuXmv, and
industry tool X complete all {\tt UARTt} tasks in under a second, while our
tool takes a few minutes due to overhead from the sampling, learning, and
SMT-check steps, making us slower on trivial model-checking problems.

\paragraph{Learning vs.~Checking Time} Figure~\ref{fig:experiments}c
illustrates the time split between learning the neural network---which
involves dataset generation and training---and verifying it as a valid
ranking function.  The lower line indicates learning time; the upper line
represents total time, with the gap showing the time spent on SMT checking. 
Extensive sampling across a broad range of trajectories covering most edge
cases led our method to learn the network directly without needing
retraining due to counterexamples in the SMT-check phase, except in four
tasks.  The plot shows that \qty{93}{\percent} of tasks were trained
successfully, generally within five minutes, and remarkably, the
\qty{70}{\percent} were completed in under a minute.  For tasks that did not
train to zero loss, the \qty{5}{\hour} time limit was not fully utilised;
the loss function stabilised at local minima in just a few minutes. 
Moreover, training was faster than verification on \qty{97}{\percent}
tasks—10 times faster on \qty{46}{\percent} and 100 times faster on
\qty{6}{\percent}.

\paragraph{Limitations} The primary limitation of our approach arises from the extended SMT-check times and the risk of getting trapped in local minima. Despite these challenges, our method consistently outperforms traditional symbolic model checkers while relying on off-the-shelf SMT solvers and machine learning optimisers. 
Additionally, our neural architecture requires numerical inputs at the word level, which limits its application to bit-level netlists. This limitation is not high-impact, as modern formal verification tools predominantly utilise Verilog RTL rather than netlist representations. 

\paragraph{Threats to Validity} The experimental results may not generalise
to other workloads.  As any work that relies on benchmarks, our benchmarks
may not be representative for other workloads.  We mitigate this threat by
selecting extremely common hardware design patterns from the standard
literature.
We remark that our data sets we use to train the neural nets do not suffer
from the common threat of training data bias, and the common
out-of-distribution problem: we train our neural net from scratch for each
benchmark using randomly generated trajectories, and do not use any
pretraining.

\section{Related Work}

Formal verification, temporal logic and model checking have been developed
for more than fifty years; key contributors have been recognised with the
1996, 2007 and 2013 ACM Turing Awards.  Here,
we restrict our discussion to algorithms that are the basis of the model checkers for SystemVerilog that available to hardware engineers today as well as on the techniques that underpin this work. 

Temporal logic describes the intended behaviour of systems and SystemVerilog Assertions---which is based on LTL---is a widely adopted language for this purpose~\cite{DBLP:conf/focs/Pnueli77,wilson}. Any temporal specifications are compositions of safety and liveness properties, where the former indicate the dangerous conditions to be avoided and the latter indicate the desirable conditions to be attained~\cite{DBLP:journals/tse/Lamport77, DBLP:journals/dc/AlpernS87}. 
Safety properties are a fragment of LTL, and can be checked using BDDs by forward fixed-point iterations~\cite{DBLP:conf/cav/KupfermanV99,DBLP:conf/birthday/ClarkeKV10, DBLP:conf/lpar/AndrausLS08}. 
Bounded model checking uses SAT and scales much better than BDDs~\cite{DBLP:conf/tacas/BiereCCZ99}, but it is only complete when the bound reaches an often unrealistically large completeness threshold~\cite{DBLP:conf/vmcai/KroeningS03}. SAT-based unbounded safety checking uses sophisticated Craig Interpolation and IC3 algorithms~\cite{DBLP:conf/cav/McMillan02,DBLP:conf/vmcai/Bradley11,DBLP:journals/corr/abs-2105-09169}.

Our work uses a one-step bounded model checking query to check the ranking function (see~Eq.~\eqref{eq:verif}), and goes beyond safety. Liveness checking for branching-time CTL is straightforward to implement using BDD-based fixed points~\cite{DBLP:conf/icalp/EmersonC80,mc2nd}. 
Our method does not support CTL; this is considered acceptable given the prevailing use of LTL-based property languages in industry. LTL model checking is commonly reduced to the fair emptiness problem and, for this purpose, bounded model checking has been generalised to $k$-liveness~\cite{DBLP:conf/fmcad/ClaessenS12,DBLP:conf/fmcad/IvriiNB18}, 
IC3 has been augmented with strongly connected components~\cite{DBLP:conf/fmcad/BradleySHZ11}, 
and BDD-based algorithms with the Emerson-Lei fixed-point computation~\cite{DBLP:conf/fmcad/BradleySHZ11, DBLP:conf/popl/EmersonL85}.
Iterative symbolic computation is the bottleneck on systems with word-level arithmetic. This is usually addressed by either computing succinct explicit-state abstractions of the system~\cite{DBLP:conf/cav/ClarkeGJLV00,DBLP:conf/cav/AbateGS24}, or by computing proof certificates based on inductive invariants and ranking functions. 

Ranking functions were introduced for termination analysis of
software~\cite{Floyd67}, and subsequently generalised to liveness
verification~\cite{DBLP:conf/icalp/LehmannPS81,
DBLP:journals/iandc/GrumbergFR85, DBLP:journals/apal/Vardi91,
DBLP:conf/popl/CookGPRV07, DBLP:conf/tacas/DimitrovaFHM16,
DBLP:conf/tacas/CookKP15, DBLP:conf/cav/AbateGR24}.  Software and hardware
model checking share common questions~\cite{DBLP:conf/isvlsi/MukherjeeKM15,
DBLP:conf/date/MukherjeeSKM16, DBLP:conf/cav/DietschHLP15}.  Early symbolic
approaches for software analysis based on constraint solving are limited to
linear ranking functions~\cite{DBLP:conf/cav/BradleyMS05,
DBLP:conf/vmcai/PodelskiR04}.  As we illustrate in
Figure~\ref{fig:illustrativeEx}, even simple examples often require
non-linear ranking functions.  These include piecewise-defined
functions~\cite{DBLP:conf/sas/Urban13, DBLP:conf/tacas/Urban15,
DBLP:conf/cav/KuraUH20}, word-level arithmetic
functions~\cite{DBLP:journals/toplas/ChenDKSW18, DBLP:conf/esop/DavidKL15},
lexicographic ranking functions~\cite{DBLP:journals/corr/LeikeH15,
DBLP:conf/icalp/BradleyMS05}, and disjoint well-founded
relations~\cite{DBLP:conf/pldi/CookPR06, DBLP:conf/tacas/CookSZ13,
DBLP:conf/cav/KroeningSTW10, DBLP:conf/lics/PodelskiR04}, and similar proof
certificates based on liveness-to-safety translation to reason about the
transitive closure of the system~\cite{DBLP:journals/toplas/PodelskiR07,
DBLP:journals/entcs/BiereAS02, DBLP:conf/hybrid/MuraliTZ24}.

Our method follows a much more lightweight approach than the symbolic
approaches above, by training ranking functions from synthetic
executions~\cite{DBLP:conf/sigsoft/Nori013}.  Deep learning has been
successfully applied to generate software and hardware designs, but without
delivering any formal guarantees~\cite{DBLP:journals/corr/abs-2107-03374,
DBLP:conf/iccad/LiuPKR23,nature-chip-layout}.  In our work, we use neural
networks {\em to represent} formal proof certificates, rather than {\em to
generate} proofs or designs.  This goes along the lines of recent work on
neural certificates, previously applied to
control~\cite{DBLP:conf/nips/ChangRG19, DBLP:journals/csysl/AbateAGP21,
DBLP:conf/hybrid/AbateAEGP21, DBLP:conf/hybrid/ZhaoZC020,
DBLP:conf/corl/DawsonQGF21, DBLP:conf/iclr/QinZCCF21,
DBLP:conf/aaai/LechnerZCH22, DBLP:conf/nips/ZikelicLVCH23,
DBLP:conf/aaai/ZikelicLHC23, DBLP:conf/aaai/NadaliM0024, quantitativecav24},
formal verification of software and probabilistic
programs~\cite{DBLP:conf/sigsoft/GiacobbeKP22, DBLP:conf/cav/AbateGR20,
DBLP:conf/concur/AbateEGPR23}, and this work applies them for the first time
to hardware model checking.

\section{Conclusion}

We have introduced a method that leverages (quantised) neural networks as
representations of ranking functions for fair termination, which we train
from synthetic executions of the system without using any external
information other than the design at hand and its specification.  We have
applied our new method to model checking SystemVerilog Assertions and
compared its performance with the state of the art on a range of
SystemVerilog designs.  We employed off-the-shelf SMT solving (Bitwuzla) and
bounded model checking (EBMC) to formally verify our neural ranking
functions~\cite{DBLP:conf/isvlsi/MukherjeeKM15,DBLP:conf/cav/NiemetzP23};
although this phase takes the majority of our compute time, with a
straightforward implementation and using tiny feed-forward neural networks,
we obtained scalability superior to traditional symbolic model checking. 
Whether alternative neural architectures as well as specialised solvers for
quantised neural networks can further improve our approach is topic of
future work~\cite{DBLP:conf/ijcai/LambGGPAV20, DBLP:conf/tacas/Amir0BK21,
DBLP:journals/tcad/MistrySB22, DBLP:journals/tcad/MatosFBMSC24}.

This is the first successful application of neural certificates to model
checking temporal logic, and introduces hardware model checking as a new
application domain for this technology.  Neural networks could be used in
many other ways to improve model checking.  Our work creates a baseline for
further development in this field and positively contributes to the safety
assurance of systems.

\section*{Acknowledgements}

We thank Matthew Leeke, Sonia Marin, and Mark Ryan for their feedback and
the anonymous reviewers for their comments and suggestions on this
manuscript.  This work was supported in part by the Advanced Research +
Invention Agency (ARIA) under the Safeguarded AI programme.

\bibliography{references}
\bibliographystyle{abbrvnat}

%\newpage
\appendix
\section{Details of the SMT Encoding of Quantised Neural Networks}
\label{sec:apd_quantNN}

The \( k^\text{th} \) hidden layer in our network comprises a fully
connected layer followed by a clamp operation that restricts outputs to the
range \([0, u]\).  This layer has \( h_k \) neurons, and the previous  layer
contains \( h_{k-1} \) neurons.  Each neuron \( i \) in the \( k^\text{th}
\) layer is defined by:
\begin{equation}
\begin{aligned}
x_i^{(k)} &= \clamp(y_i^{(k)}; u), \quad
y_i^{(k)} = b_i^{(k)} + z_i^{(k)}, \quad
z_i^{(k)} = \sum_{j=1}^{h_{k-1}} w_{ij}^{(k-1)} x_j^{(k-1)}
\end{aligned}
\label{eq:quant1}
\end{equation}

To facilitate SMT-checking modulo Bit-Vector theory, we quantise the
floating-point weights \( w_{ij} \) and biases \( b_i \) by multiplying them
by \( 2^f \) and truncating decimals, where \( f \) determines the
precision.  We define:
\[
\tilde{w}_{ij}^{(k)} = \text{trunc}(w_{ij}^{(k)} \cdot 2^f), \quad \tilde{b}^{(k)}_i = \text{trunc}(b_i^{(k)} \cdot 2^f)
\]

This transformation converts weights from floating-point values in \([0,
u]\) to integers in \([0, 2^f u]\).  To ensure consistency between
bit-vector and floating-point arithmetic, the output of each bit-vector
encoded component should be equivalent to multiplying the floating-point
output by \(2^f\) and truncating the decimals.  To achieve this, the SMT
constraints on the bit-vectors are formulated as follows:
\begin{equation}
\begin{aligned}
\bigwedge_{i=1}^{h_{k}} \left( 
\tilde{x}_i^{(k)} = \clamp(\tilde{y}_i^{(k)}; 2^f u) \land
\tilde{y}_i^{(k)} = \tilde{b}_i^{(k)} + \text{ashr}(\tilde{z}_i^{(k)}; f) \land
\tilde{z}_i^{(k)} = \sum_{j=1}^{h_{k-1}} \tilde{w}_{ij}^{(k-1)} \tilde{x}_j^{(k-1)} 
\right)
\end{aligned}
\label{eq:quant2}
\end{equation}

Here, \( \tilde{w}_{ij}^{(k-1)} \) and \( \tilde{x}_j^{(k-1)} \) are
integers in \([0, 2^f u]\), thus their product remains within \([0, 2^{2f}
u^2]\).  The sum \( \tilde{z}_i^{(k)} \) aggregates \( h_k \) such products,
resulting in \([0, 2^{2f} u^2 h_k]\).  An arithmetic right shift by \( f \)
bits scales \( \tilde{z}_i^{(k)} \) to \([0, 2^{f} u^2 h_k]\) to align with
\( \tilde{b}_i \) in \([0, 2^f u]\) (in floating-point arithmetic, the
addition would involve values in \([0, u^2 h_k]\) and \([0, u]\)).  The
clamp operation then restricts \( \tilde{y}_i^{(k)} \) to \([0, 2^f u]\),
ensuring consistency with the floating-point arithmetic, where the value
would lie within \([0, u]\).

To prevent overflow in the SMT query, we set bit-vector sizes appropriately. 
Let \( B \) be such that \( 2^B \geq 2^f u \).  Each product \(
\tilde{w}_{ij}^{(k)} \tilde{x}_j^{(k)} \) requires up to \( 2B \) bits, and
summing \( h_k \) terms necessitates additional $\log h_k$ bits.

This encoding is standard in post-training quantisation of fully connected
layers~\cite{DBLP:conf/tacas/GiacobbeHL20}.  For element-wise multiplication
layers, where each input is multiplied by a corresponding weight, we
quantise $w_i \cdot x_i$ as $\text{ashr}(\tilde{w}_i \cdot \tilde{x}_i; f)$:
Again, \( \tilde{w}_i \tilde{x}_i \) lies within \([0, 2^{2f} u^2]\), and the right shift scales it back to \([0, 2^f u^2]\), ensuring consistency with the floating point encoding.

To address the significant slowdown caused by negative numbers in the
Bitwuzla SMT-solver during our experiments, we restructured the dot product
computation in equation \ref{eq:quant2}.  By decomposing the weight vector $
\tilde{w}_{ij}$ into two non-negative components—$\tilde{w}_{ij}^{+}$
containing positive weights and $\tilde{w}_{ij}^{-}$ containing the absolute
values of negative weights—we expressed the linear layers as
\begin{equation}
\begin{aligned}
\sum_{j=1}^{h} \tilde{w}_{ij} \tilde{x}_j = \sum_{j=1}^{h} \tilde{x}_j \tilde{w}^+_{ij} - \sum_{j=1}\tilde{x}_j \tilde{w}^-_{ij} 
\end{aligned}
\label{eq:quant5}
\end{equation}

This transformation simplified multiplications to involve only non-negative
numbers and consolidated negative operations into a single subtraction,
speeding up the SMT-check in our experiments.

We further rewrite the SMT encoding—originally involving several \(
\tilde{a} \cdot \tilde{x} \) multiplications, where \( \tilde{x} \) is a
neuron value and \( \tilde{a} \) is a quantised integer weight—by replacing
these multiplications with additions and left shifts.  By factorising \(
\tilde{a} \) as a sum of powers of two, \( \tilde{a} = \sum_{i=0}^{d} c_i
\cdot 2^i \), where \( c_i \in \{0, 1\} \), the multiplication can be
rewritten as:
\[
\tilde{a} \cdot \tilde{x} = \sum_{i=0}^{d} c_i \cdot {\text{shl}(\tilde{x}; i)},
\]
where \( \text{shl}(\tilde{x}; i) \) represents left-shifting \( x \) by \(
i \) bits, effectively multiplying \( x \) by \( 2^i \).

\section{Details of the Case Studies}
\label{sec:benchmarks}

We consider ten hardware designs in our study.  These serve as benchmarks to
demonstrate the scalability of our method compared to conventional symbolic
model checkers.  They are designed to be parameterizable.

The  {\tt DELAY} models generates a positive signal \signal{sig} after a
fixed delay determined by the counter \signal{cnt}, includes a reset input
event that sets \signal{cnt} to $0$, and aims to ensure that \signal{sig}
occurs infinitely often under the assumption that the reset event
\signal{rst} is received finitely many times, resulting in the specification
$\mbox{\sf FG}~\texttt{!rst} \rightarrow \mbox{\sf GF}~\texttt{sig}$.  We
further verify $\mbox{\sf FG}~\texttt{!rst} \rightarrow \mbox{\sf
GF}~(\texttt{sig} \wedge \mbox{\sf X}~\texttt{!sig})$, to ensure
\signal{sig} doesn't remain triggered forever.

The \texttt{LCD Controller (LCD)} performs a display initialisation setup,
then awaits the \texttt{lcd\_enable} signal to transition from
\texttt{ready} to \texttt{send} for data transmission, and returns to
\texttt{ready} after a fixed interval, ensuring $\mbox{\sf
FG}~\texttt{lcd\_enable} \rightarrow \mbox{\sf GF}~\texttt{ready}$.

Similarly, {\tt Thermocouple (Tmcp.)}~transitions through stages,
\texttt{start}, \texttt{get\_data} and \texttt{pause} with suitable delay in
between, processing SPI transactions and managing transitions based on bus
activity, adhering to the specification $\mbox{\sf
FG}~\texttt{!rst}\rightarrow \mbox{\sf GF}~\texttt{get\_data}$.

The {\tt 7-Segment (7-Seg)} model alternates between two displays, ensuring
each is activated regularly unless reset, as specified by $\mbox{\sf
FG}~\texttt{!rst}\rightarrow (\mbox{\sf GF}~\texttt{disp = 0}~\wedge~
\mbox{\sf GF}~\texttt{disp = 1})$, we also verify a simpler specification
$\mbox{\sf FG}~\texttt{!rst}\rightarrow \mbox{\sf GF}~\texttt{disp = 1}$.

The {\tt i2c Stretch (i2cS)} generates timing signals \texttt{scl\_clk} and
\texttt{data\_clk} based on the ratio of input and bus clock
frequencies~\cite{i2cspec, subero2018usart}.  It monitors \texttt{rst} and
detects the \texttt{ena} signal to manage clock stretching, ensuring
$\mbox{\sf FG}~(\texttt{!rst}\,\&\,\texttt{ena}) \rightarrow \mbox{\sf
GF}~\texttt{stretch}$.

The {\tt Pulse Width Modulation (PWM)} system utilises an $N$-bit counter to
adjust pulse widths dynamically based on input, verifying the low setting of
pulse infinitely often as $\mbox{\sf GF}~\texttt{!pulse}$
\cite{holmes2003pulse}.

The {\tt VGA Controller (VGA)} manages a display interface using horizontal
and vertical counters for pixel coordinates, ensuring smooth rendering by
adjusting sync pulses and the display enable signal {\tt disp\_ena}, here we
confirm $\mbox{\sf FG}~\texttt{!rst} \rightarrow \mbox{\sf
GF}~\texttt{disp\_ena}$.

The {\tt UART Transmitter (UARTt)} toggles between {\tt wait} for preparing
data and {\tt transmit} for sending data, based on {\tt tx\_ena} requests
and {\tt clk} signals, validated by $\mbox{\sf FG}~\texttt{!rst} \rightarrow
\mbox{\sf GF}~\texttt{wait}$ ~\cite{subero2018usart}.

The {\tt Load-Store (LS)} toggles between {\tt load} and {\tt store} with a
delay implemented by counter which counts from 0 up to N when {\tt load}
then switch to {\tt store} counting back down to 0, before switching back to
{\tt load}, {\tt sig} signals a switch from {\tt load} to {\tt store}, and
we verify $\mbox{\sf FG}~\texttt{!rst} \rightarrow \mbox{\sf
GF}~\texttt{sig}$.

Lastly, the {\tt Gray Counter (Gray)} counts in Gray codes to minimise
transition errors by ensuring single bit changes between consecutive counts,
with $\mbox{\sf FG}~\texttt{!rst} \rightarrow \mbox{\sf GF}~\texttt{sig}$,
indicating regular signalling of complete cycles~\cite{weisstein2003gray}. 
Similar to the Delay module, we aim to ensure that the signal sig does not
remain triggered indefinitely.  We establish this with two distinct
specifications $\mbox{\sf FG}\texttt{!rst} \rightarrow \mbox{\sf
GF}(\texttt{sig} \wedge \mbox{\sf X}~\texttt{!sig})$ and $\mbox{\sf
FG}\texttt{!rst} \rightarrow (\mbox{\sf GF}\texttt{sig} \wedge \mbox{\sf
GF}~\texttt{!sig})$.

\begin{table}[]
    \centering
    \setlength{\tabcolsep}{5pt} % Default value: 6pt
    \renewcommand{\arraystretch}{1.4} % Adjust vertical space between rows
    \begin{tabular}{|c|c|c|}

    \hline
    Model & LTL Specification & Key Table \ref{tab:all_exp}\\
    \hline
    \multirow{2}{*}{{\tt DELAY}} 
                       & $\mbox{\sf FG}~\texttt{!rst} \rightarrow \mbox{\sf GF}~\texttt{sig}$ & Da\\ \cline{2-3}
                       &  $\mbox{\sf FG}~\texttt{!rst} \rightarrow \mbox{\sf GF}~(\texttt{sig} \wedge \mbox{\sf X}~\texttt{!sig})$ & Db \\
    \hline

    \texttt{LCD Controller} & $\mbox{\sf FG}~\texttt{lcd\_enable} \rightarrow \mbox{\sf GF}~\texttt{ready}$ & L \\
    \hline
    {\tt Thermocouple} & $\mbox{\sf FG}~\texttt{!rst}\rightarrow \mbox{\sf GF}~\texttt{get\_data}$ & T \\
    \hline
    \multirow{2}{*}{ {\tt 7-Segment}} 
                       & $\mbox{\sf FG}~\texttt{!rst}\rightarrow \mbox{\sf GF}~\texttt{disp = 1}$ & 7a\\ \cline{2-3}
                       &  $\mbox{\sf FG}~\texttt{!rst}\rightarrow (\mbox{\sf GF}~\texttt{disp = 0}~\wedge~ \mbox{\sf GF}~\texttt{disp = 1})$ & 7b \\
    \hline
    {\tt i2c Stretch} & $\mbox{\sf FG}~(\texttt{!rst}\,\&\,\texttt{ena}) \rightarrow \mbox{\sf GF}~\texttt{stretch}$& I \\
    \hline
    {\tt Pulse Width Modulation} & $\mbox{\sf GF}~\texttt{!pulse}$& P \\
    \hline
    {\tt VGA Controller} & $\mbox{\sf FG}~\texttt{!rst} \rightarrow \mbox{\sf GF}~\texttt{disp\_ena}$ & V \\
    \hline
    {\tt UART Transmitter} & $\mbox{\sf FG}~\texttt{!rst} \rightarrow \mbox{\sf GF}~\texttt{wait}$ & U \\
    \hline
    {\tt Load-Store} & $\mbox{\sf FG}~\texttt{!rst} \rightarrow \mbox{\sf GF}~\texttt{sig}$  & Ls \\
    \hline
    \multirow{3}{*}{{\tt Gray Counter}} & $\mbox{\sf FG}~\texttt{!rst} \rightarrow \mbox{\sf GF}~\texttt{sig}$ & Ga\\ \cline{2-3}
                      & $\mbox{\sf FG}\texttt{!rst} \rightarrow \mbox{\sf GF}(\texttt{sig} \wedge \mbox{\sf X}~\texttt{!sig})$ & Gb \\ \cline{2-3}
                      & $\mbox{\sf FG}\texttt{!rst} \rightarrow (\mbox{\sf GF}\texttt{sig} \wedge \mbox{\sf GF}~\texttt{!sig})$ & Gc \\
    \hline
\end{tabular}
\vspace{0.4cm}
    \caption{Model Name and LTL Specification in our Benchmark}
    \label{tab:model_spec}
\end{table}

\section{Details of the Experimental Results}
\label{sec:all_exp}

Table \ref{tab:all_exp} provides the runtimes for each tool on the 194
verification tasks considered in Section~\ref{sec:result}.  These tasks
involve verifying each hardware design across an increasing state space,
labelled numerically.  The ``Train Time'' column indicates the training
duration for the neural network in seconds, while the other columns
represent the total runtime for each tool, with the fastest tool time in
bold and the rest in grey.  In this table, our method uses the configuration
described in Section~\ref{sec:result}, with two hidden layers containing 8
and 5 neurons, respectively.  Some of our runtimes are marked with an
asterisk (*), indicating that in those cases we obtained counterexamples
using the SMT solver; these were used for retraining and then validating the
trained network.  The reported time includes all SMT checks and training. 
Table \ref{tab:case_study} summaries these results by showing the number of
tasks successfully completed by each tool for each design.  Tasks not marked
as \emph{out of time (oot.)} or \emph{did not train (dnt.)} are considered
successful.  Table \ref{tab:all_exp} serves as the basis for computing all
statistical observations discussed in Section~\ref{sec:intro} and
Section~\ref{sec:result}, except those related to the "our best" line in
Figure~\ref{fig:experiments}a.  All other components of
Figure~\ref{fig:experiments} are derived from this table.  By aggregating
the duration of each experiment in the table, including OOT instances
counted as 5 hours per experiment, the total time amounts to 104 days and 11
hours.

\section{Ablation Study}

The network architecture described in Section~\ref{sec:neuralnetwork}
includes an element-wise multiplication layer and separate trainable
parameters associated with each state of the automaton $\mathcal{A}_{\lnot
\Phi}$.  For most of our experiments in Section~\ref{sec:result} and all
experiments in Appendix~\ref{sec:all_exp}, we employ a fully connected
multilayer perceptron component with two hidden layers containing 8 and 5
neurons, respectively.  To experimentally justify our architecture, we
perform an ablation study and report the runtimes for different
configurations in Table~\ref{tab:all_exp2}.  We consider three
configurations for the two hidden layers: containing (3, 2) neurons, (5, 3)
neurons, and (15, 8) neurons, respectively.  We further replace the
element-wise multiplication layer with a fully connected layer of the same
size, denoted as `ExtL' for the extra layer.  Additionally, we explore
providing the global trainable parameters $\theta$ to all automaton states
of the automaton $\mathcal{A}_{\lnot \Phi}$, leading to a monolithic neural
ranking function $V(r,q) \equiv \bar{V}(r,q; \theta)$, where the automaton
state $q$ is given as an additional input, which we denote as `Mono'.

Given the large number of possible combinations of these modifications, we
restrict our ablation study to switching only a single configuration at a
time.  In Table~\ref{tab:all_exp2}, the column labelled `Default' contains
the results for our original configuration—the runtimes in this column are
the same as those under `Our (8, 5)' in Table~\ref{tab:all_exp}.  Following
that, we have one column for each of the three hidden layer configurations,
followed by columns for the extra layer (`ExtL'), and the monolithic neural
ranking function (`Mono').  The `our best' line in
Figure~\ref{fig:experiments}a is obtained by selecting the minimum runtime
from the `Default' and the three hidden layer configuration columns for each
of the 194 tasks.

From Table~\ref{tab:all_exp2}, we observe that our default configuration
succeeds in more cases than the alternative configurations, justifying our
choices experimentally.  Specifically, the default configuration completes
\qty{93}{\percent} of the tasks, while the three configurations with hidden
layers containing (3, 2), (5, 3), and (15, 8) neurons complete
\qty{25}{\percent}, \qty{63}{\percent}, and \qty{74}{\percent} of the tasks,
respectively.  The extra-layer configuration and the monolithic neural
ranking function complete \qty{24}{\percent}, and \qty{39}{\percent}, of the
tasks, respectively.

Generally—but not always—when a smaller network succeeds, its runtime is
lower than that of the default network.  Specifically, among the tasks
completed by the (3, 2) neuron configuration, it was faster than the default
configuration in \qty{57}{\percent}of cases; for the (5, 3) neuron
configuration, this statistic rises to 94\%.  Interestingly, this trend does
not hold when comparing the (3, 2) and (5, 3) configurations: despite having
more neurons, the (5, 3) configuration was faster than the (3, 2)
configuration in \qty{56}{\percent} of tasks.  The default configuration not
only completes more tasks than the (15, 8) configuration but is also faster
on \qty{97}{\percent} of the tasks successfully completed by the (15, 8)
configuration.  Notably, among the hidden layer configurations only the (15,
8) configuration succeeds on any of the tasks for the {\tt VGA} design,
labelled as 'V' in the table.  In \qty{67}{\percent} of the tasks that the
`Ext.~L' configuration completes, it is faster than the default
configuration; this figure rises to \qty{86}{\percent} for the `Mono'
configuration.  While the monolithic neural ranking function (`Mono') fails
on 61\% of tasks, it surprisingly succeeds on nine out of the ten tasks for
the {\tt VGA} design.  Overall, only 5 of the 194 tasks fail under all configurations in the ablation study.

\begin{table}[h!]
\centering % Center the table
\caption{Runtime comparison with the state of the art on individual tasks.}
\small
\label{tab:all_exp}

\begin{minipage}{0.5\textwidth}
    {\tiny
    \renewcommand{\arraystretch}{0.65}
    \begin{tabular}{|l|c|c|c|c|c|c|c|}
    \hline
    \multirow{2}{*}{Tasks} & {Train} & \multicolumn{5}{c|}{Total Time per Tool (in sec.)} \\\cline{3-7}
              & Time  & our (8,5)  & nuXmv & ABC & X & Y \\\hline\hline
     Da$_1   $ & 6   & \gt{44  } & \Bf{2.5 }  & \gt{398 }  & \gt{442} & \gt{oot.}   \\\hline
     Da$_2   $ & 10  & \gt{51  } & \Bf{7   }  & \gt{1759}  & \gt{802} & \gt{oot.} \\\hline
     Da$_3   $ & 7   & \gt{80  } & \Bf{29  }  & \gt{8666}  & \gt{801} & \gt{oot.} \\\hline
     Da$_4   $ & 7   & \gt{92  } & \Bf{121 }  & \gt{oot.}  & \gt{815} & \gt{oot.} \\\hline
     Da$_5   $ & 41  & \Bf{157 } & \gt{292 }  & \gt{oot.}  & \gt{788} & \gt{oot.}  \\\hline
     Da$_6   $ & 24  & \Bf{162 } & \gt{529 }  & \gt{oot.}  & \gt{814} & \gt{oot.} \\\hline
     Da$_7   $ & 15  & \Bf{197 } & \gt{870 }  & \gt{oot.}  & \gt{809} & \gt{oot.} \\\hline
     Da$_8   $ & 36  & \Bf{214 } & \gt{1277}  & \gt{oot.}  & \gt{793} & \gt{oot.} \\\hline
     Da$_9   $ & 23  & \Bf{321 } & \gt{1809}  & \gt{oot.}  & \gt{809} & \gt{oot.}  \\\hline
     Da$_{10}$ & 15  & \Bf{306 } & \gt{2448}  & \gt{oot.}  & \gt{804} & \gt{oot.} \\\hline
     Da$_{11}$ & 44  & \Bf{390 } & \gt{3516}  & \gt{oot.}  & \gt{789} & \gt{oot.} \\\hline
     Da$_{12}$ & 17  & \Bf{412 } & \gt{4461}  & \gt{oot.}  & \gt{788} & \gt{oot.} \\\hline
     Da$_{13}$ & 22  & \Bf{674 } & \gt{oot.}  & \gt{oot.}  & \gt{808} & \gt{oot.}  \\\hline
     Da$_{14}$ & 26  & \gt{1500} & \gt{oot.}  & \gt{oot.}  & \Bf{815} & \gt{oot.} \\\hline
     Da$_{15}$ & 45  & \gt{3365} & \gt{oot.}  & \gt{oot.}  & \Bf{802} & \gt{oot.} \\\hline
     Da$_{16}$ & 124 & \gt{8684} & \gt{oot.}  & \gt{oot.}  & \Bf{813} & \gt{oot.} \\\hline\hline

     P$_1$    &  2   & \gt{30  }  & \Bf{1.5  } &  \gt{926 } & \gt{792} & \gt{  oot.}   \\\hline
     P$_2$    &  1   & \gt{26  }  & \Bf{6.5  } &  \gt{5087} & \gt{776} & \gt{  oot.}   \\\hline
     P$_3$    &  4   & \gt{97  }  & \Bf{30   } &  \gt{oot.} & \gt{785} & \gt{  oot.}   \\\hline
     P$_4$    &  12  & \Bf{69  }  & \gt{137  } &  \gt{oot.} & \gt{782} & \gt{  oot.}   \\\hline
     P$_5$    &  14  & \Bf{94  }  & \gt{638  } &  \gt{oot.} & \gt{788} & \gt{  oot.}    \\\hline
     P$_6$    &  30  & \Bf{177 }  & \gt{2563 } &  \gt{oot.} & \gt{773} & \gt{  oot.}   \\\hline
     P$_7$    &  64  & \Bf{365 }  & \gt{10667} &  \gt{oot.} & \gt{787} & \gt{  oot.}   \\\hline
     P$_8$    &  334 & \gt{1028}  & \gt{ oot.} &  \gt{oot.} & \Bf{798} & \gt{  oot.}   \\\hline
     P$_9$    & 17 & \gt{2611}  & \gt{ oot.} &  \gt{oot.} &   \Bf{781} & \gt{  oot.}    \\\hline
     P$_{10}$ & 20 & \gt{6527}  & \gt{ oot.}  & \gt{oot.} &   \Bf{788} & \gt{oot.  }   \\\hline
     P$_{11}$ & 33 & \gt{9353}  & \gt{ oot.}  & \gt{oot.} & \Bf{787} & \gt{oot.  }   \\\hline 
     P$_{12}$ & dnt. & \gt{ -  }  & \gt{ oot.}  & \gt{oot.} & \Bf{785} & \gt{oot.  }   \\\hline \hline

     L$_{1} $  &  4    & \gt{42  } & \Bf{0.8  } & \gt{129 } & \gt{55 }  &  \gt{oot. } \\\hline
     L$_{2} $  &  5    & \gt{63  } & \Bf{1.9  } & \gt{1189} & \gt{215}   & \gt{ oot.}  \\\hline
     L$_{3} $  &  5    & \gt{53  } & \Bf{12   } & \gt{1712} & \gt{808}   & \gt{ oot.}  \\\hline
     L$_{4} $  &  8    & \gt{83  } & \Bf{12   } & \gt{oot.} & \gt{799}   & \gt{ oot.}  \\\hline
     L$_{5} $  &  46   & \Bf{245 } & \gt{951  } & \gt{oot.} & \gt{838}   & \gt{ oot.}  \\\hline
     L$_{6} $  &  18   & \Bf{297 } & \gt{4444 } & \gt{oot.} & \gt{815}   & \gt{ oot.}  \\\hline
     L$_{7} $  &  22   & \Bf{360 } & \gt{3262 } & \gt{oot.} & \gt{843}   & \gt{ oot.}  \\\hline
     L$_{8} $  &  31   & \Bf{335 } & \gt{11061} & \gt{oot.} & \gt{828}   & \gt{ oot.}  \\\hline
     L$_{9} $  &  38   & \Bf{355 } & \gt{1743 } & \gt{oot.} & \gt{807}   & \gt{ oot.}  \\\hline
     L$_{10}$  &  32   & \Bf{470 } & \gt{oot. } & \gt{oot.} & \gt{837}   & \gt{ oot.}  \\\hline
     L$_{11}$  &  23   & \Bf{360 } & \gt{oot. } & \gt{oot.} & \gt{825}   & \gt{ oot.}  \\\hline
     L$_{12}$  &  102  & \Bf{622 } & \gt{oot. } & \gt{oot.} & \gt{805}   & \gt{ oot.}  \\\hline
     L$_{13}$  &  79   & \gt{3321} & \gt{oot. } & \gt{oot.} & \Bf{850}   & \gt{ oot.}  \\\hline
     L$_{14}$  &  181 & \gt{7133} & \gt{oot. } & \gt{oot.} &  \Bf{817}   & \gt{ oot.}  \\\hline\hline

     I$_1 $   & 6     & \gt{49 }   & \Bf{2.5  } & \gt{201 }  & \gt{169 } &  \gt{oot.  }    \\\hline
     I$_2 $   & 9     & \gt{74 }   & \Bf{10   } & \gt{1195}  & \gt{oot.}  & \gt{ oot. }     \\\hline
     I$_3 $   & 12    & \gt{102}   & \Bf{44   } & \gt{6396}  & \gt{793 } &  \gt{oot.  }    \\\hline
     I$_4 $   & 78    & \Bf{163}   & \gt{196  } & \gt{oot.}  & \gt{801 } &  \gt{oot.  }    \\\hline
     I$_5 $   & 42    & \Bf{170}   & \gt{527  } & \gt{oot.}  & \gt{795 } &  \gt{oot.  }    \\\hline
     I$_6 $   & 25    & \Bf{173}   & \gt{1038 } & \gt{oot.}  & \gt{797 } &  \gt{oot.  }    \\\hline
     I$_7 $   & 29    & \Bf{173}   & \gt{1801 } & \gt{oot.}  & \gt{794 } &  \gt{oot.  }    \\\hline
     I$_8 $   & 62    & \Bf{245}   & \gt{2738 } & \gt{oot.}  & \gt{800 } &  \gt{oot.  }    \\\hline
     I$_9 $   & 49    & \Bf{289}   & \gt{9288 } & \gt{oot.}  & \gt{797 } &  \gt{oot.  }    \\\hline
     I$_{10}$ & 63    & \Bf{474}   & \gt{15674} & \gt{oot.}  & \gt{oot.}  & \gt{  oot.}     \\\hline
     I$_{11}$ & 62    & \Bf{354}   & \gt{oot. } & \gt{oot.}  & \gt{807 } &  \gt{ oot. }    \\\hline
     I$_{12}$ & 97    & \Bf{386}   & \gt{oot. } & \gt{oot.}  & \gt{805 } &  \gt{ oot. }    \\\hline
     I$_{13}$ & 134   & \Bf{358}   & \gt{oot. } & \gt{oot.}  & \gt{798 } &  \gt{ oot. }    \\\hline
     I$_{14}$ & 114   & \Bf{417}   & \gt{oot. } & \gt{oot.}  & \gt{818 } &  \gt{ oot. }    \\\hline
     I$_{15}$ & 342   & \Bf{661}   & \gt{oot. } & \gt{oot.}  & \gt{792 } &  \gt{ oot. }    \\\hline
     I$_{16}$ & 140  &  \Bf{585}.  & \gt{oot. } & \gt{oot.}  & \gt{812 } &  \gt{ oot. }   \\\hline
     I$_{17}$ & 332  &  \Bf{615}.  & \gt{oot. } & \gt{oot.}  & \gt{798 } &  \gt{ oot. }    \\\hline
     I$_{18}$ & 474  &  \gt{908}.  & \gt{oot. } & \gt{oot.}  & \Bf{824 } &  \gt{ oot. }    \\\hline
     I$_{19}$ & oot.  & \gt{oot.}.  & \gt{oot. } & \gt{oot.}  & \Bf{817 } &  \gt{ oot. }    \\\hline
     I$_{20}$ & oot.  & \gt{oot.}.  & \gt{oot. } & \gt{oot.}  & \Bf{811 } &  \gt{ oot. }    \\\hline\hline

     Ga$_1$      & 3    & \gt{18  }  & \Bf{0.3  } & \gt{53  } & \gt{81 }   & \gt{oot.} \\\hline
     Ga$_2$      & 3    & \gt{30  }  & \Bf{1.2  } & \gt{78  } & \gt{293}   & \gt{oot.} \\\hline
     Ga$_3$      & 6    & \gt{25  }  & \Bf{5    } & \gt{233 } & \gt{784}   & \gt{oot.} \\\hline
     Ga$_4$      & 10   & \gt{37  }  & \Bf{22   } & \gt{6490} & \gt{795}   & \gt{oot.} \\\hline
     Ga$_5$      & 4    & \Bf{62  }  & \gt{96   } & \gt{6217} & \gt{789}   & \gt{oot.} \\\hline
     Ga$_6$      & 12   & \Bf{102 }  & \gt{447  } & \gt{oot.} & \gt{786}   & \gt{oot.} \\\hline
     Ga$_7$      & 17   & \Bf{175 }  & \gt{2062 } & \gt{oot.} & \gt{801}   & \gt{oot.} \\\hline
     Ga$_8$      & 28   & \Bf{299 }  & \gt{12935} & \gt{oot.} & \gt{797}   & \gt{oot.} \\\hline
     Ga$_9$      & 39   & \Bf{639 }  & \gt{oot. } & \gt{oot.} & \gt{811}   & \gt{oot.} \\\hline
     Ga$_{10}$   & 118  & \gt{1566}  & \gt{oot. } & \gt{oot.} & \Bf{792}   & \gt{oot.} \\\hline
     Ga$_{11}$   & 218  & \gt{5790}  & \gt{oot. } & \gt{oot.} & \Bf{787}   & \gt{oot.} \\\hline \hline
    
     Db$_1   $ & 12   & \gt{89  } & \Bf{ 3   }  & \gt{231 }  & \gt{993}  & \gt{oot.}   \\\hline
     Db$_2   $ & 10   & \gt{85  } & \Bf{ 7   }  & \gt{379 }  & \gt{6568} & \gt{oot.} \\\hline
     Db$_3   $ & 12   & \gt{181 } & \Bf{ 30  }  & \gt{3396}  & \gt{oot.} & \gt{oot.} \\\hline
     Db$_4   $ & 14   & \gt{204 } & \Bf{ 130 }  & \gt{oot.}  & \gt{oot.} & \gt{oot.} \\\hline
     Db$_5   $ & 30   & \Bf{312 } & \gt{ 308 }  & \gt{oot.}  & \gt{4931} & \gt{oot.}  \\\hline
     Db$_6   $ & 145  & \Bf{471 } & \gt{ 570 }  & \gt{oot.}  & \gt{oot.} & \gt{oot.} \\\hline
     Db$_7   $ & 158  & \Bf{711 } & \gt{ 917 }  & \gt{oot.}  & \gt{oot.} & \gt{oot.} \\\hline
     Db$_8   $ & 170  & \Bf{532 } & \gt{ 1349}  & \gt{oot.}  & \gt{oot.} & \gt{oot.} \\\hline
     Db$_9   $ & 25   & \Bf{662 } & \gt{ 1912}  & \gt{oot.}  & \gt{oot.} & \gt{oot.}  \\\hline
     Db$_{10}$ & 214  & \Bf{746 } & \gt{ 2605}  & \gt{oot.}  & \gt{oot.} & \gt{oot.} \\\hline
     Db$_{11}$ & 226  & \Bf{885 } & \gt{ 3597}  & \gt{oot.}  & \gt{oot.} & \gt{oot.} \\\hline
     Db$_{12}$ & 200  & \Bf{930 } & \gt{ 4439}  & \gt{oot.}  & \gt{oot.} & \gt{oot.} \\\hline
     Db$_{13}$ & 363  & \Bf{2654} & \gt{oot. }  & \gt{oot.}  & \gt{oot.} & \gt{oot.}  \\\hline
     Db$_{14}$ & 728 & \Bf{3893} & \gt{oot. }  & \gt{oot.}  & \gt{oot.} & \gt{oot.} \\\hline
     Db$_{15}$ & 588  & \Bf{5700} & \gt{oot. }  & \gt{oot.}  & \gt{oot.} & \gt{oot.} \\\hline
     Db$_{15}$ & 797 & \gt{12697} & \gt{oot. }  & \gt{oot.}  & \gt{oot.} & \gt{oot.} \\\hline \hline

     Ls$_1   $ & 6    & \gt{51 }   &  \Bf{16   } & \gt{768  } & \gt{510} & \gt{oot.}   \\\hline
     Ls$_2   $ & 5    & \Bf{53 }   &  \gt{56   } & \gt{10772} & \gt{539} & \gt{oot.} \\\hline
     Ls$_3   $ & 3    & \Bf{78 }   &  \gt{251  } & \gt{oot. } & \gt{580} & \gt{oot.} \\\hline
     Ls$_4   $ & 19   & \Bf{126}   &  \gt{1263 } & \gt{oot. } & \gt{621} & \gt{oot.} \\\hline
     Ls$_5   $ & 21   & \Bf{185}   &  \gt{2612 } & \gt{oot. } & \gt{633} & \gt{oot.}  \\\hline
     Ls$_6   $ & 26   & \Bf{218}   &  \gt{6722 } & \gt{oot. } & \gt{662} & \gt{oot.} \\\hline
     Ls$_7   $ & 22   & \Bf{403}   &  \gt{9490 } & \gt{oot. } & \gt{668} & \gt{oot.} \\\hline
     Ls$_8   $ & 24   & \Bf{300}   &  \gt{12665} & \gt{oot. } & \gt{674} & \gt{oot.} \\\hline

    \end{tabular}
    }
\end{minipage}%  This comment sign % here makes sure there is no space between the two minipages
\begin{minipage}{0.5\textwidth}
    {\tiny
    \renewcommand{\arraystretch}{0.65}
    \begin{tabular}{|l|c|c|c|c|c|c|c|}
    
    \hline
    \multirow{2}{*}{Tasks} & {Train} & \multicolumn{5}{c|}{Total Time per Tool (in sec.)} \\\cline{3-7}
              & Time & our (8,5)  & nuXmv  & ABC & X & Y \\\hline\hline
     7a$_1$    &  5    &  \gt{21  } & \Bf{2    } & \gt{39  } & \gt{28 }   & \gt{ oot.}   \\\hline
     7a$_2$    &  5    &  \Bf{34  } & \Bf{8    } & \gt{119 } & \gt{192}   & \gt{ oot.}   \\\hline
     7a$_3$    &  4    &  \Bf{24  } & \gt{70   } & \gt{614 } & \gt{467}   & \gt{ oot.}   \\\hline
     7a$_4$    &  5    &  \Bf{38  } & \gt{1405 } & \gt{1469} & \gt{680}   & \gt{ oot.}   \\\hline
     7a$_5$    &  5    &  \Bf{47  } & \gt{11605} & \gt{oot.} & \gt{812}   & \gt{ oot.}   \\\hline
     7a$_6$    &  4    &  \Bf{58  } & \gt{oot. } & \gt{oot.} & \gt{806}   & \gt{ oot.}   \\\hline
     7a$_7$    &  5    &  \Bf{86  } & \gt{oot. } & \gt{oot.} & \gt{820}   & \gt{ oot.}   \\\hline
     7a$_8$    &  6    &  \Bf{104 } & \gt{oot. } & \gt{oot.} & \gt{815}   & \gt{ oot.}   \\\hline
     7a$_9$    &  15   &  \Bf{215 } & \gt{oot. } & \gt{oot.} & \gt{816}   & \gt{ oot.}   \\\hline
     7a$_{10}$ &  10   &  \Bf{154 } & \gt{oot. } & \gt{oot.} & \gt{816}   & \gt{oot. }   \\\hline
     7a$_{11}$ &  11   &  \Bf{242 } & \gt{oot. } & \gt{oot.} & \gt{817}   & \gt{oot. }   \\\hline
     7a$_{12}$ &  20   &  \Bf{208 } & \gt{oot. } & \gt{oot.} & \gt{817}   & \gt{oot. }  \\\hline 
     7a$_{13}$ &  18   &  \Bf{495 } & \gt{oot. } & \gt{oot.} & \gt{815}   & \gt{oot. }   \\\hline
     7a$_{14}$ &  26   &  \gt{977 } & \gt{oot. } & \gt{oot.} & \Bf{819}   & \gt{oot. }   \\\hline
     7a$_{15}$ &  68   &  \gt{2077} & \gt{oot. } & \gt{oot.} & \Bf{824}   & \gt{oot. }   \\\hline \hline

     T$_{1} $ &  7    & \gt{22  }  & \Bf{ 0.6  } & \gt{ 2   } & \gt{1}   & \gt{oot.} \\\hline
     T$_{2} $ &  23   & \gt{67  }  & \Bf{ 9    } & \gt{ 62  } & \gt{470}   & \gt{oot.} \\\hline
     T$_{3} $ &  11   & \Bf{95  }  & \gt{ 361  } & \gt{ 234 } & \gt{41}   & \gt{oot.} \\\hline
     T$_{4} $ &  11   & \Bf{98  }  & \gt{ 601  } & \gt{ 344 } & \gt{103}   & \gt{oot.} \\\hline
     T$_{5} $ &  16   & \Bf{164 }  & \gt{ 306  } & \gt{ 1872} & \gt{414}   & \gt{oot.} \\\hline
     T$_{6} $ &  12   & \Bf{156 }  & \gt{ 573  } & \gt{ 1246} & \gt{246}   & \gt{oot.} \\\hline
     T$_{7} $ &  12   & \Bf{182 }  & \gt{ 1192 } & \gt{ 2195} & \gt{308}   & \gt{oot.} \\\hline
     T$_{8} $ &  23   & \Bf{210 }  & \gt{ 1935 } & \gt{ oot.} & \gt{532}   & \gt{oot.} \\\hline
     T$_{9} $ &  17   & \Bf{326 }  & \gt{ 4224 } & \gt{ oot.} & \gt{798}   & \gt{oot.} \\\hline
     T$_{10} $&  20   & \Bf{516 }  & \gt{ 6365 } & \gt{ oot.} & \gt{796}   & \gt{oot.} \\\hline
     T$_{11} $&  44   & \Bf{389 }  & \gt{ 9691 } & \gt{ oot.} & \gt{797}   & \gt{oot.} \\\hline
     T$_{12} $&  39   & \gt{ 932}  & \gt{ 15129} & \gt{ oot.} & \Bf{791}   & \gt{oot.} \\\hline
     T$_{13} $&  39   & \gt{ 949}  & \gt{ oot. } & \gt{ oot.} & \Bf{807}   & \gt{oot.} \\\hline
     T$_{14} $&  104  & \gt{1552}  & \gt{ oot. } & \gt{ oot.} & \Bf{803}   & \gt{oot.} \\\hline
     T$_{15} $&  79 & \gt{6020} &  \gt{oot.  }&  \gt{oot. } &   \Bf{805}   & \gt{oot.} \\\hline
     T$_{16} $&  60 & \gt{7331}  & \gt{ oot. } & \gt{ oot.} &   \Bf{800}   & \gt{oot.} \\\hline
     T$_{17} $&  118 & \gt{13042}  & \gt{ oot. } & \gt{ oot.} & \Bf{786}   & \gt{oot.} \\\hline\hline

     V$_1 $    & dnt. &  \gt{-}   &  \Bf{25  }  & \gt{26  } &  \gt{90 }  & \gt{  oot.}    \\\hline
     V$_2 $    & dnt. &  \gt{-}   &  \Bf{781 }  & \gt{oot.} &  \gt{794}  & \gt{  oot.}    \\\hline
     V$_3 $    & dnt. &  \gt{-}   &  \gt{4448}  & \gt{2870} &  \Bf{796}  & \gt{oot.  }    \\\hline
     V$_4 $    & dnt. &  \gt{-}   &  \gt{oot.}  & \gt{4748} &  \Bf{795}  & \gt{oot.  }    \\\hline
     V$_5 $    & dnt. &  \gt{-}   &  \gt{oot.}  & \gt{oot.} &  \Bf{792}  & \gt{oot.  }    \\\hline
     V$_6 $    & dnt. &  \gt{-}   &  \gt{oot.}  & \gt{oot.} &  \Bf{792}  & \gt{oot.  }    \\\hline
     V$_7 $    & dnt. &  \gt{-}   &  \gt{oot.}  & \gt{oot.} &  \Bf{793}  & \gt{  oot.}    \\\hline
     V$_8 $    & dnt. &  \gt{-}   &  \gt{oot.}  & \gt{oot.} &  \Bf{801}  & \gt{  oot.}    \\\hline
     V$_9 $    & dnt. &  \gt{-}   &  \gt{oot.}  & \gt{oot.} &  \Bf{805}  & \gt{  oot.}    \\\hline
     V$_{10} $ & dnt. &  \gt{-}   &  \gt{oot.}  & \gt{oot.} &  \Bf{841}  & \gt{oot.  }    \\\hline \hline

     U$_1$   & 7     & \gt{35  } & \Bf{0.04}  & \gt{0.48} & \gt{1.34} & \gt{oot.} \\\hline
     U$_2$   & 11    & \gt{29  } & \Bf{0.06}  & \gt{0.39} & \gt{1.3 } & \gt{oot.} \\\hline
     U$_3$   & 19    & \gt{148 } & \Bf{0.08}  & \gt{0.45} & \gt{1.94} & \gt{oot.} \\\hline
     U$_4$   & 53    & \gt{74 }  & \Bf{0.4 }  & \gt{0.45} & \gt{1.21} & \gt{oot.} \\\hline
     U$_5$   & 103   & \gt{188}  & \Bf{0.24}  & \gt{0.42} & \gt{1.22} & \gt{oot.} \\\hline
     U$_6$   & 295   & \gt{371}  & \Bf{0.09}  & \gt{0.39} & \gt{1.25} & \gt{oot.} \\\hline
     U$_7$   & 26    & \gt{222 } & \Bf{0.14}  & \gt{0.49} & \gt{1.32} & \gt{oot.} \\\hline
     U$_8$   & 514.  & \gt{1804} & \Bf{0.1 }  & \gt{0.46} & \gt{1.21} & \gt{oot.} \\\hline
     U$_9$   & 51    & \gt{567}  & \Bf{0.12}  & \gt{0.72} & \gt{1.29} & \gt{oot.} \\\hline
     U$_{10}$& 46    & \gt{112 } & \gt{3.28}  & \Bf{0.79} & \gt{1.41} & \gt{oot.} \\\hline \hline

     Gb$_1$      &  3   & \gt{41  }  & \Bf{0.3  } & \gt{49  } & \gt{824  }&  \gt{oot. }\\\hline
     Gb$_2$      &  3   & \gt{97  }  & \Bf{1    } & \gt{196 } & \gt{417  }&  \gt{oot. }\\\hline
     Gb$_3$      &  3   & \gt{160 }  & \Bf{5    } & \gt{358 } & \gt{3204 } & \gt{ oot.} \\\hline
     Gb$_4$      &  3   & \gt{207 }  & \Bf{24   } & \gt{1559} & \gt{3661 } & \gt{ oot.} \\\hline
     Gb$_5$      &  5   & \gt{302 }  & \Bf{110  } & \gt{oot.} & \gt{13164} &  \gt{ oot.}. \\\hline
     Gb$_6$      &  9   & \Bf{292 }  & \gt{511  } & \gt{oot.} & \gt{oot. } & \gt{ oot.} \\\hline
     Gb$_7$      &  7   & \Bf{862 }  & \gt{2441 } & \gt{oot.} & \gt{3341 } & \gt{ oot.} \\\hline
     Gb$_8$      &  8   & \Bf{2958}  & \gt{14518} & \gt{oot.} & \gt{oot. } & \gt{ oot.} \\\hline
     Gb$_9$      &  10  & \Bf{3847}  & \gt{oot. } & \gt{oot.} & \gt{oot. } & \gt{ oot.} \\\hline
     Gb$_{10}$   &  18  & \Bf{4676}  & \gt{oot. } & \gt{oot.} & \gt{oot. } & \gt{ oot.} \\\hline
     Gb$_{11}$   &  36  & \Bf{8834}  & \gt{oot. } & \gt{oot.} & \gt{oot. } & \gt{ oot.} \\\hline \hline

     Gc$_1$      &  8    & \gt{41   } & \Bf{0.3  } &  \gt{88   } & \gt{94  }  &  \gt{oot.} \\\hline
     Gc$_2$      &  12   & \gt{52   } & \Bf{1.25 } &  \gt{139  } & \gt{539  } &  \gt{oot.} \\\hline
     Gc$_3$      &  5    & \gt{100  } & \Bf{5    } &  \gt{4428 } & \gt{3349 }  & \gt{oot.} \\\hline
     Gc$_4$      &  6    & \gt{132  } & \Bf{24   } &  \gt{4373 } & \gt{3688 }  & \gt{oot.} \\\hline
     Gc$_5$      &  73   & \gt{260  } & \Bf{105  } &  \gt{oot. } & \gt{oot. }  & \gt{oot.} \\\hline
     Gc$_6$      &  17   & \Bf{256  } & \gt{491  } &  \gt{oot. } & \gt{3488 }  & \gt{oot.} \\\hline
     Gc$_7$      &  50   & \Bf{1091 } & \gt{2387 } &  \gt{oot. } & \gt{oot. } &  \gt{oot.} \\\hline
     Gc$_8$      &  176  & \Bf{947  } & \gt{14287} &  \gt{oot. } & \gt{oot. } &  \gt{oot.} \\\hline
     Gc$_9$      &  580  & \Bf{2300 } & \gt{oot. } &  \gt{oot. } & \gt{oot. } &  \gt{oot.} \\\hline
     Gc$_{10}$   &  1685  &\gt{5052}  &\gt{ oot.}  & \gt{ oot.}  &\gt{ oot.} &  \gt{oot.} \\\hline
     Gc$_{11}$   &  169  &\gt{9888}  &\gt{ oot.}  & \gt{ oot.}  &\gt{ oot.} &  \gt{oot.} \\\hline \hline

     7b$_1$    &  5    & \gt{ 30  } & \Bf{1.5  } & \gt{ 69  } &  \gt{558 }  &  \gt{oot. }  \\\hline
     7b$_2$    &  6    & \gt{ 60  } & \Bf{5    } & \gt{ 350 } &  \gt{5352}   & \gt{ oot.}   \\\hline
     7b$_3$    &  5    & \gt{ 59  } & \Bf{20   } & \gt{ 3606} &  \gt{oot.}   & \gt{ oot.}   \\\hline
     7b$_4$    &  5    & \Bf{ 75  } & \gt{1463 } & \gt{ 2663} &  \gt{2332}   & \gt{ oot.}   \\\hline
     7b$_5$    &  5    & \Bf{ 102 } & \gt{13208} & \gt{ oot.} &  \gt{oot.}   & \gt{ oot.}   \\\hline
     7b$_6$    &  6    & \Bf{ 125 } & \gt{oot. } & \gt{ oot.} &  \gt{oot.}   & \gt{ oot.}   \\\hline
     7b$_7$    &  7    & \Bf{ 181 } & \gt{oot. } & \gt{ oot.} &  \gt{oot.}   & \gt{ oot.}   \\\hline
     7b$_8$    &  12   & \Bf{ 207 } & \gt{oot. } & \gt{ oot.} &  \gt{oot.}   & \gt{ oot.}   \\\hline
     7b$_9$    &  9    & \Bf{ 438 } & \gt{oot. } & \gt{ oot.} &  \gt{oot.}   & \gt{ oot.}   \\\hline
     7b$_{10}$ &  14   & \Bf{ 238 } & \gt{oot. } & \gt{ oot.} &  \gt{oot.}   & \gt{oot. }   \\\hline
     7b$_{11}$ &  15   & \Bf{ 439$^{*}$} & \gt{oot. } & \gt{oot. } &  \gt{oot.}   & \gt{oot. }   \\\hline
     7b$_{12}$ &  14   & \Bf{ 343 } & \gt{oot. } & \gt{ oot.} &  \gt{oot.}   & \gt{oot. }  \\\hline 
     7b$_{13}$ &  22   & \Bf{ 578 } & \gt{oot. } & \gt{ oot.} &  \gt{oot.}   & \gt{oot. }   \\\hline
     7b$_{14}$ &  65   & \Bf{2121$^{*}$} & \gt{oot. } & \gt{ oot.} &  \gt{oot.}   & \gt{oot. }   \\\hline
     7b$_{15}$ &  48   & \Bf{ 2187} & \gt{oot. } & \gt{ oot.} &  \gt{oot.}   & \gt{oot. }   \\\hline \hline

     Ls$_9   $ &  24  & \Bf{473   } &  \gt{oot.}  & \gt{oot.}  & \gt{709} & \gt{oot.}  \\\hline
     Ls$_{10}$ &  19  & \Bf{486   } &  \gt{oot.}  & \gt{oot.}  & \gt{703} & \gt{oot.} \\\hline
     Ls$_{11}$ &  22  & \Bf{558   } &  \gt{oot.}  & \gt{oot.}  & \gt{727} & \gt{oot.} \\\hline
     Ls$_{12}$ &  75  & \Bf{695   } &  \gt{oot.}  & \gt{oot.}  & \gt{709} & \gt{oot.} \\\hline
     Ls$_{13}$ &  22  & \gt{1420  } &  \gt{oot.}  & \gt{oot.}  & \Bf{750} & \gt{oot.}  \\\hline
     Ls$_{14}$ &  125 & \gt{4336  } &  \gt{oot.}  & \gt{oot.}  & \Bf{791} & \gt{oot.} \\\hline
     Ls$_{15}$ &  197 & \gt{14533$^{*}$} &  \gt{oot.}  & \gt{oot.}  & \Bf{832} & \gt{oot.} \\\hline
     Ls$_{15}$ &  88  & \gt{oot.$^{*}$}  &  \gt{oot.}  & \gt{oot.}  & \Bf{873} & \gt{oot.} \\\hline

    \end{tabular} 
    }
\end{minipage}

\end{table}

\begin{table}[ht]
\centering % Center the table
\caption{Ablation Study Runtime.}
\small
\label{tab:all_exp2}
\begin{minipage}{0.5\textwidth}
    {\tiny
    \renewcommand{\arraystretch}{0.65}
    \begin{tabular}{|l|c|c|c|c|c|c|c|}
    \hline

    \multirow{2}{*}{Tasks} & \multicolumn{6}{c|}{Total Time per Setup (in sec.)} \\\cline{2-7}
             & Default & (3, 2) & (5, 3) & (15, 8) & Ext.L  & Mono \\\hline\hline
     Da$_1   $ & 44   &  14   & 21    & 121  &  29    &   44        \\\hline
     Da$_2   $ & 51   &  11   & 23    & 172  &  30    &   66      \\\hline
     Da$_3   $ & 80   &  15   & 27    & 324  &  64    &   48      \\\hline
     Da$_4   $ & 92   &  21   & 37    & 330  &  71    &   53     \\\hline
     Da$_5   $ & 157  &  47   & 41    & 609  &  81    &   57       \\\hline
     Da$_6   $ & 162  &  90   & 52    & 410  &  84    &   fail      \\\hline
     Da$_7   $ & 197  &  169  & 119   & 427  &  84    &   64      \\\hline
     Da$_8   $ & 214  &  182  & 109   & 595  &  71    &   fail      \\\hline
     Da$_9   $ & 321  &  221  & 104   & 789  &  115   &   160       \\\hline
     Da$_{10}$ & 306  &  255  & 132   & 654  &  125   &   fail      \\\hline
     Da$_{11}$ & 390  &  420  & 161   & 1093 &  192   &   fail      \\\hline
     Da$_{12}$ & 412  &  383  & 177   & 1200 &  135   &   fail      \\\hline
     Da$_{13}$ & 674  &  443  & 204   & 1446 &  fail  &   375       \\\hline
     Da$_{14}$ & 1500 &  686  & 618   & 3463 &  fail  &   643      \\\hline
     Da$_{15}$ & 3365 &  1239 & 2217  & 6525 &  fail  &   fail      \\\hline
     Da$_{16}$ & 8684 &  4526 & 3272  & fail &  fail  &   1597      \\\hline\hline

     P$_1$    & 30   &  65    &   15    &  83   &  18   &  18     \\\hline
     P$_2$    & 26   &  77    &   12    &  109  &  16   &  85      \\\hline
     P$_3$    & 97   &  72    &   14    &  255  &  44   &  25      \\\hline
     P$_4$    & 69   &  85    &   20    &  fail &  89   &  71      \\\hline
     P$_5$    & 94   &  601   &   24    &  fail &  102  &  41       \\\hline
     P$_6$    & 177  &  374   &   68    &  fail &  147  &  119     \\\hline
     P$_7$    & 365  &  695   &   154   &  fail &  241  &  248     \\\hline
     P$_8$    & 1028 &  1364  &   fail  &  fail &  462  &  355     \\\hline
     P$_9$    & 2611 &  6030  &   fail  &  fail &  1561 &  fail      \\\hline
     P$_{10}$ & 6527 &  5667  &   fail  &  fail &  1597 &  1943     \\\hline
     P$_{11}$ & 9353 &  29992 &   fail  &  fail &  8019 &  8190     \\\hline 
     P$_{12}$ & fail &  fail  &   fail  &  fail &  fail &  fail     \\\hline \hline

     L$_{1} $  & 42    &  fail & 15    & 172  &  fail  &   30      \\\hline
     L$_{2} $  & 63    &  fail & 23    & 349  &  fail  &   29      \\\hline
     L$_{3} $  & 53    &  fail & 34    & 335  &  fail  &   fail      \\\hline
     L$_{4} $  & 83    &  fail & 32    & 575  &  fail  &   fail      \\\hline
     L$_{5} $  & 245   &  fail & 51    & 1043 &  fail  &   173      \\\hline
     L$_{6} $  & 297   &  fail & 116   & 829  &  fail  &   fail      \\\hline
     L$_{7} $  & 360   &  fail & 234   & 1106 &  fail  &   fail      \\\hline
     L$_{8} $  & 335   &  fail & fail  & 1329 &  fail  &   fail      \\\hline
     L$_{9} $  & 355   &  fail & fail  & 1328 &  fail  &   fail      \\\hline
     L$_{10}$  & 470   &  fail & 297   & 2751 &  fail  &   fail      \\\hline
     L$_{11}$  & 360   &  fail & 429   & 1679 &  fail  &   179      \\\hline
     L$_{12}$  & 622   &  fail & 224   & 2660 &  fail  &   fail      \\\hline
     L$_{13}$  & 3321  &  fail & 1905  & 7537 &  fail  &   929      \\\hline
     L$_{14}$  & 7133  &  fail & 3041  & fail &  fail  &   fail      \\\hline\hline

     I$_1 $   & 49    & fail   &  fail   & fail  & fail  &   56         \\\hline
     I$_2 $   & 74    & fail   &  fail   & 288   & fail  &   73         \\\hline
     I$_3 $   & 102   & fail   &  fail   & fail  & fail  &   368        \\\hline
     I$_4 $   & 121   & fail   &  921    & fail  & fail  &   fail       \\\hline
     I$_5 $   & 170   & fail   &  fail   & 596   & 254   &   fail       \\\hline
     I$_6 $   & 173   & fail   &  fail   & fail  & fail  &   fail       \\\hline
     I$_7 $   & 173   & fail   &  fail   & 987   & fail  &   fail       \\\hline
     I$_8 $   & 245   & fail   &  fail   & fail  & fail  &   fail       \\\hline
     I$_9 $   & 289   & fail   &  fail   & 815   & fail  &   fail       \\\hline
     I$_{10}$ & 474   & fail   &  fail   & 1637  & fail  &   fail       \\\hline
     I$_{11}$ & 354   & fail   &  fail   & 1595  & fail  &   fail       \\\hline
     I$_{12}$ & 386   & fail   &  fail   & 1016  & fail  &   fail       \\\hline
     I$_{13}$ & 358   & fail   &  fail   & fail  & 754   &   fail       \\\hline
     I$_{14}$ & 417   & fail   &  fail   & 1116  & fail  &   fail       \\\hline
     I$_{15}$ & 661   & fail   &  fail   & fail  & fail  &   fail       \\\hline
     I$_{16}$ & 585   & fail   &  fail   & 1499  & fail  &   fail      \\\hline
     I$_{17}$ & 615   & fail   &  fail   & fail  & fail  &   fail       \\\hline
     I$_{18}$ & 908   & fail   &  fail   & 2372  & fail  &   fail       \\\hline
     I$_{19}$ & fail  & fail   &  fail   & fail  & fail  &   fail       \\\hline
     I$_{20}$ & fail  & fail   &  fail   & fail  & fail  &   fail       \\\hline\hline

     Ga$_1$      & 18    &   5    &   9    & 66    &  22    &  8      \\\hline
     Ga$_2$      & 30    &   5    &   12   & 91    &  14    &  18      \\\hline
     Ga$_3$      & 25    &   9    &   19   & 148   &  30    &  25      \\\hline
     Ga$_4$      & 37    &   12   &   27   & 346   &  47    &  15      \\\hline
     Ga$_5$      & 62    &   14   &   30   & 425   &  56    &  46      \\\hline
     Ga$_6$      & 102   &   39   &   40   & 309   &  73    &  81      \\\hline
     Ga$_7$      & 175   &   113  &   124  & 648   &  101   &  170      \\\hline
     Ga$_8$      & 299   &   444  &   163  & 1040  &  fail  &  178      \\\hline
     Ga$_9$      & 639   &   648  &   308  & 2178  &  fail  &  fail      \\\hline
     Ga$_{10}$   & 1566  &   1014 &   849  & 4727  &  fail  &  1395      \\\hline
     Ga$_{11}$   & 5790  &   2207 &   3318 & 10035 &  fail  &  1480      \\\hline \hline

     Db$_1   $ & 89   &  fail  &  27     & fail  &  fail &    fail    \\\hline
     Db$_2   $ & 85   &  fail  &  24     & 380   &  fail &    fail  \\\hline
     Db$_3   $ & 181  &  fail  &  50     & 1066  &  fail &    fail  \\\hline
     Db$_4   $ & 204  &  fail  &  79     & 1276  &  fail &    fail  \\\hline
     Db$_5   $ & 312  &  fail  &  187    & 2264  &  fail &    fail   \\\hline
     Db$_6   $ & 471  &  fail  &  fail   & 1889  &  fail &    fail  \\\hline
     Db$_7   $ & 711  &  fail  &  fail   & 1971  &  fail &    fail  \\\hline
     Db$_8   $ & 532  &  fail  &  fail   & fail  &  fail &    fail  \\\hline
     Db$_9   $ & 662  &  fail  &  fail   & 2706  &  fail &    fail   \\\hline
     Db$_{10}$ & 746  &  fail  &  fail   & fail  &  fail &    fail  \\\hline
     Db$_{11}$ & 885  &  fail  &  fail   & 2949  &  fail &    fail  \\\hline
     Db$_{12}$ & 930  &  fail  &  fail   & 3292  &  fail &    fail  \\\hline
     Db$_{13}$ & 2654 &  fail  &  fail   & fail  &  fail &    fail   \\\hline
     Db$_{14}$ & 3893 &  fail  &  fail   & 6326  &  fail &    fail  \\\hline
     Db$_{15}$ & 5700 &  fail  &  fail   & fail  &  fail &    fail  \\\hline
     Db$_{15}$ & 12697 &  fail  &  fail   & fail  &  fail &   fail  \\\hline

    \end{tabular}
    }
\end{minipage}%  This comment sign % here makes sure there is no space between the two minipages
\begin{minipage}{0.5\textwidth}
    {\tiny

    \renewcommand{\arraystretch}{0.65}
    \begin{tabular}{|l|c|c|c|c|c|c|}
    
    \hline
    \multirow{2}{*}{Tasks} & \multicolumn{6}{c|}{Total Time per Setup (in sec.)} \\\cline{2-7}
             & Default & (3, 2) & (5, 3) & (15, 8) & Ext.L  & Mono \\\hline\hline
              
     7a$_1$    & 21   &  16   &  fail  & 73   &  fail  &   fail      \\\hline
     7a$_2$    & 34   &  fail &  fail  & 130  &  fail  &   fail      \\\hline
     7a$_3$    & 24   &  484  &  fail  & 160  &  fail  &   fail      \\\hline
     7a$_4$    & 38   &  675  &  fail  & 189  &  fail  &   14      \\\hline
     7a$_5$    & 47   &  fail &  fail  & 199  &  fail  &   fail      \\\hline
     7a$_6$    & 58   &  210  &  fail  & 349  &  fail  &   fail      \\\hline
     7a$_7$    & 86   &  fail &  fail  & 329  &  fail  &   fail      \\\hline
     7a$_8$    & 104  &  432  &  fail  & 479  &  fail  &   118      \\\hline
     7a$_9$    & 215  &  611  &   64   & 611  &  fail  &   fail      \\\hline
     7a$_{10}$ & 154  &  542  &  fail  & 641  &  fail  &   fail      \\\hline
     7a$_{11}$ & 242  &  574  &  fail  & 858  &  fail  &   164      \\\hline
     7a$_{12}$ & 208  &  790  &  fail  & 835  &  fail  &   fail     \\\hline 
     7a$_{13}$ & 495  &  981  &  fail  & 1109 &  fail  &   399      \\\hline
     7a$_{14}$ & 977  &  988  &  fail  & 4137 &  fail  &   fail      \\\hline
     7a$_{15}$ & 2077 &  1579 &  fail  & 7811 &  fail  &   fail      \\\hline \hline

     T$_{1} $ & 22     &  fail   &  12    &  70   &  fail  &   16      \\\hline
     T$_{2} $ & 67     &  fail   &  30    &  221  &  fail  &   fail      \\\hline
     T$_{3} $ & 95     &  fail   &  46    &  235  &  fail  &   42      \\\hline
     T$_{4} $ & 98     &  fail   &  45    &  364  &  fail  &   54      \\\hline
     T$_{5} $ & 164    &  fail   &  59    &  269  &  fail  &   96      \\\hline
     T$_{6} $ & 156    &  fail   &  76    &  316  &  fail  &   57      \\\hline
     T$_{7} $ & 182    &  fail   &  94    &  320  &  162   &   fail      \\\hline
     T$_{8} $ & 210    &  fail   &  85    &  377  &  fail  &   fail      \\\hline
     T$_{9} $ & 326    &  fail   &  330   &  1088 &  fail  &   115      \\\hline
     T$_{10} $& 516    &  fail   &  179   &  1131 &  fail  &   fail      \\\hline
     T$_{11} $& 389    &  fail   &  246   &  3382 &  fail  &   263      \\\hline
     T$_{12} $& 932    &  fail   &  522   &  3846 &  fail  &   451      \\\hline
     T$_{13} $& 949    &  fail   &  509   &  3931 &  fail  &   fail      \\\hline
     T$_{14} $& 1552   &  fail   &  730   &  4121 &  fail  &   1023      \\\hline
     T$_{15} $& 6020   &  fail   &  1613  &  fail &  fail  &   1276      \\\hline
     T$_{16} $& 7331   &  fail   &  4896  &  fail &  fail  &   10333      \\\hline
     T$_{17} $& 13042  &  fail   &  8295  &  fail &  8406  &   fail      \\\hline\hline

     V$_1 $    &  fail & fail  &  fail  & 272   & fail  &   82       \\\hline
     V$_2 $    &  fail & fail  &  fail  & fail  & fail  &   303       \\\hline
     V$_3 $    &  fail & fail  &  fail  & fail  & fail  &   273       \\\hline
     V$_4 $    &  fail & fail  &  fail  & 2292  & fail  &   637       \\\hline
     V$_5 $    &  fail & fail  &  fail  & 3927  & fail  &   948       \\\hline
     V$_6 $    &  fail & fail  &  fail  & 15612 & fail  &   1135       \\\hline
     V$_7 $    &  fail & fail  &  fail  & fail  & fail  &   2129       \\\hline
     V$_8 $    &  fail & fail  &  fail  & fail  & fail  &   3247       \\\hline
     V$_9 $    &  fail & fail  &  fail  & fail  & fail  &   13628       \\\hline
     V$_{10}$  &  fail & fail  &  fail  & fail  & fail  &   fail       \\\hline \hline

     U$_1$   & 35    &  fail  & 20  & fail &  62   &   fail      \\\hline
     U$_2$   & 29    &  fail  & 16  & 109  &  234  &   21      \\\hline
     U$_3$   & 148   &  fail  & 24  & 90   &  391  &   105      \\\hline
     U$_4$   & 69    &  fail  & 25  & 593  &  429  &   101      \\\hline
     U$_5$   & 74    &  fail  & 32  & 97   &  760  &   fail      \\\hline
     U$_6$   & 188   &  fail  & 34  & 281  &  1792 &   206      \\\hline
     U$_7$   & 222   &  fail  & 42  & 178  &  251  &   195      \\\hline
     U$_8$   & 1804  &  fail  & 81  & fail &  fail &   4677      \\\hline
     U$_9$   & 567   &  fail  & 303 & fail &  fail &   fail      \\\hline
     U$_{10}$& 112   &  fail  & 422 & fail &  fail &   fail      \\\hline \hline

     Gb$_1$      &  41    &  fail & 41    & 118   &  fail    &  fail   \\\hline
     Gb$_2$      &  97    &  fail & 40    & 268   &  fail    &  fail   \\\hline
     Gb$_3$      &  160   &  fail & 37    & 332   &  fail    &  fail   \\\hline
     Gb$_4$      &  207   &  fail & 61    & fail  &  fail    &  fail   \\\hline
     Gb$_5$      &  302   &  fail & 82    & 831   &  fail    &  fail   \\\hline
     Gb$_6$      &  292   &  fail & 162   & fail  &  fail    &  fail   \\\hline
     Gb$_7$      &  862   &  fail & 245   & fail  &  fail    &  fail   \\\hline
     Gb$_8$      &  2958  &  fail & 427   & 3201  &  fail    &  fail   \\\hline
     Gb$_9$      &  3847  &  fail & 691   & 6981  &  fail    &  fail   \\\hline
     Gb$_{10}$   &  4676  &  fail & 1500  & 17888 &  fail    &  fail   \\\hline
     Gb$_{11}$   &  8834  &  fail & 3820  & fail  &  fail    &  fail   \\\hline \hline

     Gc$_1$      &  41   &  fail  & 23     & 147  & fail  &  fail  \\\hline
     Gc$_2$      &  52   &  fail  & 21     & 156  & fail  &  fail  \\\hline
     Gc$_3$      &  100  &  fail  & 49     & 288  & fail  &  fail  \\\hline
     Gc$_4$      &  132  &  fail  & 38     & 381  & fail  &  fail  \\\hline
     Gc$_5$      &  260  &  fail  & 54     & 1600 & fail  &  fail  \\\hline
     Gc$_6$      &  256  &  fail  & 123    & 1792 & fail  &  fail  \\\hline
     Gc$_7$      &  1091 &  fail  & 228    & 2343 & fail  &  fail  \\\hline
     Gc$_8$      &  947  &  fail  & 549    & 2864 & fail  &  fail  \\\hline
     Gc$_9$      &  2300 &  fail  & 1470   & 6352 & fail  &  fail  \\\hline
     Gc$_{10}$   &  5052 &  fail  & 2155   & fail & fail  &  fail  \\\hline
     Gc$_{11}$   &  169  &  9888  & 6288   & fail & fail  &  fail  \\\hline \hline

     7b$_1$    & 30     &  fail   & 22    & 121  &  fail  &  fail   \\\hline
     7b$_2$    & 60     &  fail   & 46    & 217  &  fail  &  fail   \\\hline
     7b$_3$    & 59     &  fail   & 33    & 280  &  fail  &  fail   \\\hline
     7b$_4$    & 75     &  fail   & 39    & 398  &  fail  &  fail   \\\hline
     7b$_5$    & 102    &  fail   & 44    & 752  &  fail  &  fail   \\\hline
     7b$_6$    & 125    &  fail   & 60    & 1124 &  fail  &  fail   \\\hline
     7b$_7$    & 181    &  fail   & 117   & 996  &  fail  &  fail   \\\hline
     7b$_8$    & 207    &  fail   & 143   & 1306 &  fail  &  fail   \\\hline
     7b$_9$    & 438    &  fail   & 260   & 3085 &  fail  &  fail   \\\hline
     7b$_{10}$ & 238    &  fail   & 198   & 1839 &  fail  &  fail   \\\hline
     7b$_{11}$ & 439    &  fail   & 210   & 2349 &  fail  &  fail   \\\hline
     7b$_{12}$ & 343    &  fail   & 220   & 1907 &  fail  &  fail  \\\hline 
     7b$_{13}$ & 578    &  fail   & 366   & 3597 &  fail  &  fail   \\\hline
     7b$_{14}$ & 2121   &  fail   & 2070  & 5107 &  fail  &  fail   \\\hline
     7b$_{15}$ & 2187   &  fail   & fail  & fail &  fail  &  fail   \\\hline

    \end{tabular} 
    }
\end{minipage}

\end{table}

\label{sec:apd_ablation_study}

\end{document}